\documentclass[conference]{IEEEtran}
\usepackage{cite}
\usepackage{amsmath,amssymb,amsfonts}
\usepackage[ruled,vlined,linesnumbered]{algorithm2e}
\SetNlSty{}{}{\,.}
\usepackage[T1]{fontenc}
\usepackage{graphicx}
\usepackage{textcomp}
\usepackage[dvipsnames,table]{xcolor}
\usepackage{booktabs}
\usepackage{array}
\usepackage{multirow}
\usepackage{tabularx}
\usepackage{makecell}
\usepackage{adjustbox}
\usepackage{enumitem}
\usepackage{pifont}
\usepackage{siunitx}
\usepackage{newtxtext}
\newcolumntype{Y}{>{\raggedright\arraybackslash}X}

\usepackage{tikz}
\usetikzlibrary{shapes.geometric,arrows.meta,positioning,fit,calc,backgrounds}
\tikzset{
  block/.style      = {rectangle,draw,rounded corners,minimum width=3.8cm,minimum height=1.2cm,text centered,font=\footnotesize,fill=blue!10},
  smallblock/.style = {rectangle,draw,rounded corners,minimum width=3.6cm,minimum height=1cm,text centered,font=\scriptsize,fill=green!10},
  arrow/.style      = {thick,-{Stealth}},
  group/.style      = {draw,dashed,inner sep=0.3cm,rounded corners},
  model/.style      = {rectangle,minimum width=2cm,minimum height=1cm,text centered,draw=black,fill=green!20},
  data/.style       = {rectangle,draw=black,fill=white,text centered,minimum height=1.2cm,minimum width=1.2cm},
  decision/.style   = {diamond,draw=black,fill=orange!40,text centered,aspect=2,inner sep=1pt,minimum height=1.2cm},
  line/.style       = {draw,-latex'}
}
\newcommand*\circled[1]{%
  \tikz[baseline=(char.base)]{\node[shape=circle,fill=black,text=white,inner sep=0.6pt] (char) {#1};}}

\def\BibTeX{{\rm B\kern-.05em{\sc i\kern-.025em b}\kern-.08em
    T\kern-.1667em\lower.7ex\hbox{E}\kern-.125emX}}

\usepackage[hidelinks]{hyperref}
\usepackage{xspace}
\usepackage{cleveref}
\begin{document}

\newcommand{\system}{RAMSeS\xspace}


\setcounter{section}{0}
\setcounter{subsection}{0}
\setcounter{subsubsection}{0}
\setcounter{figure}{0}
\setcounter{table}{0}
\setcounter{equation}{0}
\setcounter{footnote}{0}
\setcounter{algocf}{0} 

\title{RAMSeS: Robust and Adaptive Model Selection for Time-Series Anomaly Detection Algorithms}

\author{
\IEEEauthorblockN{Mohamed Abdelmaksoud}
\IEEEauthorblockA{
BIFOLD \& TU Berlin \\
Berlin, Germany \\
mohamed@tu-berlin.de}
\and
\IEEEauthorblockN{Sheng Ding}
\IEEEauthorblockA{
University of Stuttgart \\
Stuttgart, Germany \\
sheng.ding@ias.uni-stuttgart.de}
\and
\IEEEauthorblockN{Andrey Morozov}
\IEEEauthorblockA{
University of Stuttgart \\
Stuttgart, Germany \\
andrey.morozov@ias.uni-stuttgart.de}
\and
\IEEEauthorblockN{Ziawasch Abedjan}
\IEEEauthorblockA{
BIFOLD \& TU Berlin \\
Berlin, Germany \\
abedjan@tu-berlin.de}
}

\maketitle

\begin{abstract}
Time-series data vary widely across domains, making a universal anomaly detector impractical. Methods that perform well on one dataset often fail to transfer because what counts as an anomaly is context dependent. The key challenge is to design a method that performs well in specific contexts while remaining adaptable across domains with varying data complexities.
We present the Robust and Adaptive Model Selection for Time-Series Anomaly Detection RAMSeS framework. RAMSeS comprises two branches: (i) a stacking ensemble optimized with a genetic algorithm to leverage complementary detectors. (ii) An adaptive model-selection branch identifies the best single detector using techniques including Thompson sampling, robustness testing with generative adversarial networks, and Monte Carlo simulations. This dual strategy exploits the collective strength of multiple models and adapts to dataset-specific characteristics.
We evaluate RAMSeS and show that it outperforms prior methods on F1.
\end{abstract}


\section{Introduction}
\label{sec:introduction}

Time-series are prevalent across domains such as finance, industry, and healthcare~\cite{keogh2005hot, wagner2023timesead, autotsad_thorsten_24, JiangLJPE20, hishida2025beyond, PaparrizosLBHEE21, LiuBPP24}. They encode complex system dynamics and often exhibit heterogeneity in seasonality, regimes, and sampling irregularities~\cite{goswami2022unsupervised, liu2025tsb}.
The diversity in time-series structures is accompanied by a wide range of anomaly types that occur in practice, such as isolated spikes, contextual anomalies, and collective anomalies~\cite{chandola2009anomaly, lindemann2021survey, boniol2021unsupervised, boniol2021sand, BoniolPKPTEF22, BoniolPP23, BoniolPP24, LiuP24a, BhattacharyyaJTW11, LiaoLLT13}. Each category captures a distinct notion of abnormality, and manifestations are domain-dependent~\cite{goldstein2016comparative, liu2025tsb}.
Structural diversity across time-series and the variety of anomaly types impede the realization of a ``one-size-fits-all'' model. Methods that excel on one series often degrade on another, and a single series can be subject to concept drifts, where the normal behavior and distribution changes over time~\cite{goldstein2016comparative, goswami2022unsupervised, liu2025tsb}. 
In clinical patient monitoring, where, physiological signals such as ECG, blood pressure, and respiratory rate differ in scale, sampling frequency, and what constitutes anomalous behavior~\cite{goswami2022unsupervised, ucr}. An irregular heartbeat, a sudden spike in blood pressure, and an abnormal breathing pattern reflect different anomaly types and therefore require different models. Moreover, even within a single patient's ECG stream, the optimal detector can shift: during rest periods, simple threshold-based methods suffice, but during exercise or stress, adaptable pattern-based detectors are needed to distinguish anomalous arrhythmia from expected elevated heart rate~\cite{wagner2023timesead, campos2021unsupervised}.

\begin{figure}[t]
\centering
\includegraphics[width=\linewidth]{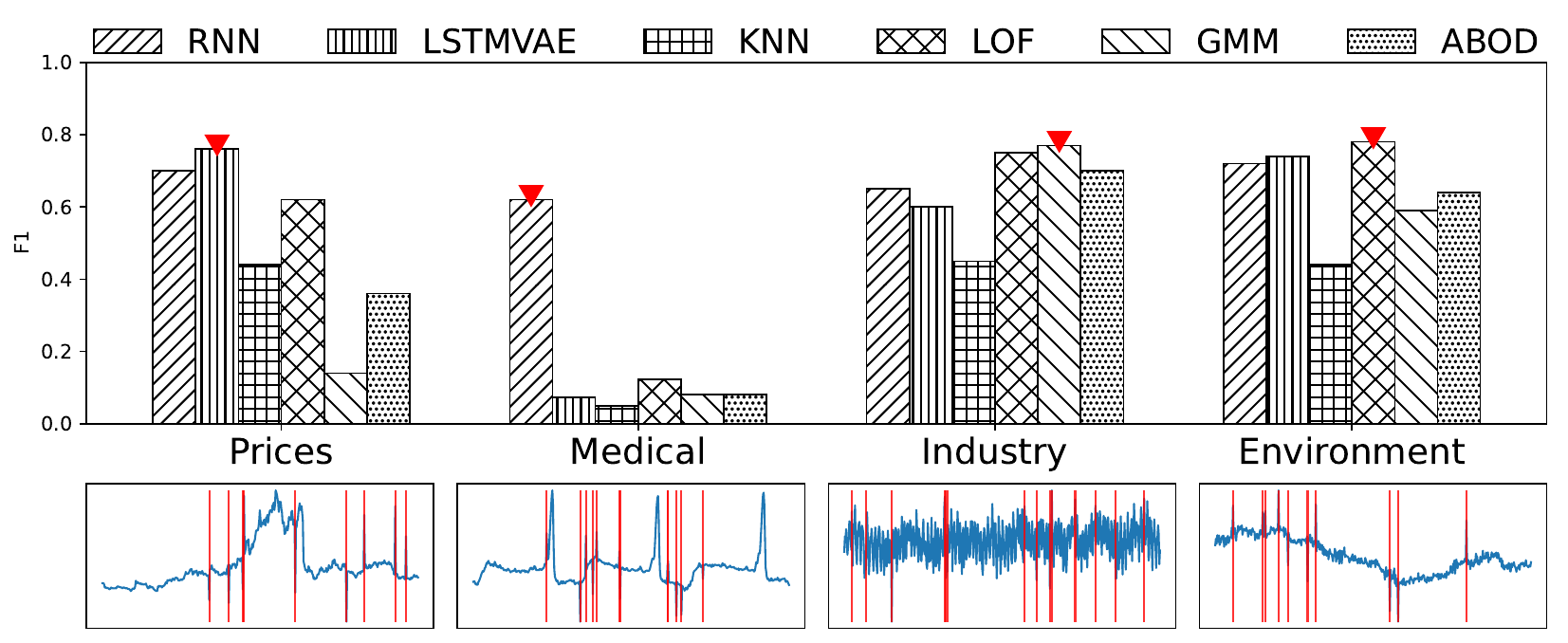}
\vspace{-6mm} 
\caption{F1 score for six anomaly detection models from statistical and neural-network families across four domains (Prices, Medical, Industry, and Environment)~\cite{ucr, skab, smd, vandenburg2020evaluation_TCPD}. Red vertical lines indicate spike anomalies. The triangle marks the best-performing model in each domain. 
}
\label{fig:f1_panels_from_files}
\vspace{-5mm}
\end{figure}

Anomaly detection in heterogeneous setups is challenging not only because the problem is multifaceted, but also because existing approaches face practical limitations and data scarcity. 
For example, distance- and density-based methods such as k-nearest neighbors and Local Outlier Factor rely on stable local neighborhoods~\cite{breunig2000lof, RamaswamyRS00_NN}, in reality the series might be subject to temporal non-stationarity, i.e., different density of data points across time intervals. Reconstruction-based approaches, such as variational autoencoders, assume anomalies yield larger reconstruction errors than normal patterns~\cite{ParkHK18_LSTMVAE, MalhotraVSA15_LSTMAD, jung2021time}. Assumptions break under shifts in scale, overexpressive autoencoders, or concept drift~\cite{goswami2022unsupervised, TAB_qiu_2025, liu2025tsb, autotsad_thorsten_24}. Figure~\ref{fig:f1_panels_from_files} depicts performance of statistical and neural-network-based models on four different time-series, each from a different domain. No approach outperforms all other on all domains confirming the aforementioned limitations.

Existing work has attempted to address these limitations either by selecting the best fit single model per time series or by aggregating multiple models in ensembles~\cite{TAB_qiu_2025, goswami2022unsupervised, liu2025tsb, autotsad_thorsten_24, Ma0ZA23_86_IOE, AggarwalS15_5_OE}. Nevertheless, significant open challenges remain. Current single-model selection algorithms are efficient and interpretable on stable segments but degrade under regime changes, non-stationarity, and borderline anomalies~\cite{liu2025tsb}. Prior ensemble techniques exploit the collective signals of multiple detectors but are highly sensitive to subset model selection and often assume comparable score scales~\cite{liu2025tsb}. 
To bridge this gap, we propose \system (\textbf{R}obust and \textbf{A}daptive \textbf{M}odel \textbf{Se}lection for Time-\textbf{S}eries Anomaly Detection). 
\system introduces two novel approaches within the same framework: a genetic-algorithm-based stacking ensemble that uses self-supervised learning to effectively search through possible ensemble candidates and
a single-model selection module that combines linear Thompson sampling with robustness testing.
In summary, we make the following contributions:

\begin{itemize}[leftmargin=0.2cm]
    \item \textit{Genetic-algorithm–based stacking ensemble (GA-Ens).} 
    Our approach combines a genetic algorithm with a meta-learner. The entanglement helps to guide the subset selection process of the genetic algorithm via scores learned by the meta-learner. As a result, our approach converges significantly faster than exhaustive grid and Bayesian search~\cite{AlibrahimL21_gen_vs_grid_bayes}. 
    
    \item \textit{Single model selection} traverses models using contextual linear Thompson sampling (LinTS) and $\varepsilon$-greedy exploration: LinTS is conditioned on window-level context and balances exploration and exploitation, yielding rankings that follow local regime changes. Simultaneously, robustness is assessed via three tests: Generative Adversarial Network (GAN)–based perturbations, Monte Carlo simulations, and an off-by-threshold sensitivity analysis.  Then, the LinTS-based rankings and robustness signals are combined via Markov-based rank aggregation to produce a stable consensus order. 
    \item We present an experimental evaluation on known benchmark datasets~\cite{ucr, smd, skab} and show that each of our proposed approaches, ensemble and model selection, significantly outperform the state-of-the-art counter parts~\cite{goswami2022unsupervised,liu2025tsb}.
\end{itemize}



\section{Related Work}
\label{sec:related_work}
There are six lines of related work: \emph{time series anomaly detection} and model selection targeting the no one-size-fits-all challenge, \emph{model selection via surrogate metrics} that avoid labels, \emph{learning-based adaptive selection} with reinforcement learning and bandits, \emph{meta-learning} for per-dataset model choice, \emph{model ensembling} for combining multiple base learners’ scores, and \emph{benchmarking frameworks}. 

\subsection{Time Series Anomaly Detection and Model Selection}
Time series anomaly detection (TSAD) spans statistical forecasting, residual analysis, density-based, and deep learning approaches including autoencoders, recurrent networks, and Transformers~\cite{hundman2018detecting, AudibertMGMZ20_USAD, TuliCJ22_TranAD,XuWWL22_AnomalyTransformer,breunig2000lof}.
Recent work has explored online detection, model-based diagnostics, and toolchain integration for cyber-physical systems~\cite{ding2019line, ding2021kraken,imbsa2020,ding2022tool,hu2024adaptive,ding2022imu}, as well as adaptive model selection and automated evaluation~\cite{ying2020automated, ding2023auto}.
However, these methods treat model evaluation as a static ranking task without incorporating robustness testing or temporal reliability. In contrast, \system integrates adaptive model selection with robustness-aware ensemble optimization, enabling consistent performance under diverse and evolving conditions.

\subsection{Surrogate Metrics for Model Selection}
\label{subsec:surrogate}
In the absence of labels, surrogate metrics estimate detector quality by combining multiple signals. Goswami et al.~\cite{goswami2022unsupervised} propose a composite score that aggregates prediction error, detector centrality, and performance on synthetic anomalies via rank aggregation to approximate supervised orderings. Although effective on curated datasets, this method assumes that synthetic anomalies resemble real deviations and may overemphasize detectors with high centrality. Jung et al.~\cite{jung2021time} present LaF-AD, which weights detectors by the variance of anomaly probabilities across bootstraps to reward stability. However, it does not explicitly test robustness to distribution shift or adversarial noise and lacks online adaptivity. \system incorporates robustness evaluations GAN-based perturbations, off-by-threshold tests, and Monte Carlo simulations, to stress-test and rank detectors under realistic conditions.

\subsection{Learning-Based Adaptive Selection}
\label{subsec:learning}
Adaptive selectors learn policies that switch between models over time. Zhang et al.~\cite{ZhangWB22} formulate model selection as a Markov decision process (MDP) over contextual features such as detector scores and inter-model agreement, training a policy with confidence-based rewards. This approach captures temporal adaptation but relies on crafted states and requires tuning and training effort. Ngo et al.~\cite{ngo2021adaptive} apply contextual bandits in hierarchical edge environments to balance accuracy and latency. Their selection is trained offline and optimizes deployment cost rather than robustness or generalization. In contrast, \system integrates an adaptive selection policy based on LinTS with $\varepsilon$-greedy, allowing contextual adaptation while preserving robustness to noise and shifting conditions.

\subsection{Meta-Learning for Time Series Model Choice}
\label{subsec:metalearning}
Meta-learning approaches infer the most suitable detector from dataset characteristics. Sylligardos et al.~\cite{sylligardos2023choose} demonstrate that dataset-specific model choice can be learned and incorporated into AutoML pipelines, but the resulting selection remains static at inference time and ignores intra-series regime changes. Boniol et al.~\cite{BoniolSPTP24} present ADecimo, a practical selector for time-series anomaly detection, which similarly focuses on dataset-level choice rather than online adaptation.
To overcome such static behavior, \system extends beyond per-dataset meta-learning by enabling continuous, context-aware model adaptation throughout the time series.

\subsection{Ensemble Selection}
\label{subsec:ensemble}
At a high level, ensembles differ along two axes: (i) \emph{scope}, aggregating scores from all detectors versus selecting a subset, and (ii) \emph{fusion rule}, how per-detector scores are combined.

Outlier Ensemble (OE) instantiates three canonical score-level rules: \emph{AVG},  mean of detector scores to reduce variance. \emph{MAX}, pointwise maximum to highlight strong outlier evidence and reduce bias, and \emph{AOM}, which averages maxima over random detector subsets to balance bias–variance~\cite{AggarwalS15_5_OE}. Iterative Outlier Ensemble (IOE) iteratively refines a pseudo ground truth by updating with the closest detector score vector until convergence~\cite{Ma0ZA23_86_IOE}. SELECT performs \emph{component selection} via two complementary schemes: a \emph{Vertical} strategy that prefers detectors agreeing on anomalous samples (model consensus), and a \emph{Horizontal} strategy that, per sample, prefers detectors that appear most reliable for that sample. The selected outputs are then ensembled using consensus techniques~\cite{RayanaA16_119_SELECT}. HITS Adapting hub/authority centrality, HITS builds a bipartite graph between detectors and data points to produce detector reliability weights~\cite{Kleinberg99_67,Ma0ZA23_86_IOE}.
Existing ensemble selection techniques remain biased toward group consensus, even when the collective decision is incorrect. \system mitigates this by employing a GA-Ens to ensure consistency and scalability.

\subsection{Benchmarking-Based Evaluation Paradigms}
TSB-UAD provides an end-to-end TSAD suite emphasizing standardized evaluation protocols and reproducibility~\cite{PaparrizosKBTPF22}. Similarly, Schmidl et al.~\cite{SchmidlWP22} conduct a comprehensive comparison of classical and deep models, revealing variance across datasets and metrics. AutoTSAD introduces a label-free framework that synthesizes regime-rich series with injected anomalies to optimize configurations and ensembles~\cite{autotsad_thorsten_24}. Recently, TSB-AutoAD evaluates selection, ensemble, and generation strategies across domains, exposing strong variability and trade-offs between accuracy and computational cost, and proposing selective ensemble as a lightweight hybrid~\cite{liu2025tsb}. These efforts demonstrate that detector performance is highly context-dependent and that a simple ensemble can be accurate but computationally expensive~\cite{SchmidlWP22, autotsad_thorsten_24, liu2025tsb}.
Building on these insights, \system incorporates an ensemble, reinforcement learning and robustness evaluations to provide systematic, dataset-independent tests for model selection.

\section{Problem Formulation and \system Overview}
\label{sec:problem-overview}
We formalize the problem of model selection for TSAD and outlines our framework, \system.

\subsection{Problem formulation}
\label{sec:problem}
We are considering problems of single model and ensemble selection for TSAD. Before defining the objective, we first introduce time series, anomaly types and model candidates.

\noindent\textit{A Time series} $X=\{x_t\}_{t=1}^T$ can be univariate ($x_t\in\mathbb{R}$) or multivariate ($x_t\in\mathbb{R}^{d_X}$). A fixed \emph{sliding window} per series is generated. Given window length $w_X$ and stride $s_X$, we obtain windows $\{W_i\}_{i=1}^{N_X}$ with $W_i=\{x_t\}_{t=t_i}^{t_i+w_X-1}$.

\noindent\textit{Anomaly types} include \emph{point}, \emph{contextual}, and \emph{collective} anomalies~\cite{goswami2022unsupervised}. Point anomalies affect a single timestamp, such as \emph{spikes} and \emph{flips}. Contextual anomalies violate local behavior, such as seasonality or local trends. Collective anomalies arise when a subsequence is abnormal as a group although individual points  appear normal~\cite{goswami2022unsupervised}.

\noindent\textit{Candidates:}
 A set of base anomaly detection models $\mathcal{D}=\{d_1,\dots,d_M\}$ serve as candidates. For each window $W_i$, a model $d_m$ outputs a normalized anomaly score $s_{i,m}\in[0,1]$. 

\noindent\textit{Objective:}
With the given time series dataset $X$ and the set of candidates $\mathcal{D}$, the optimization goals for ensemble and model selection are slightly different. The former aims at obtaining an ensemble, i.e., a subset of candidates, and the latter aims at finding the best candidate both, with the goal of achieving high detection quality, measured by common scores, such as the harmonic mean (F1) of precision and recall or the area under the curve AUC-PR, or a combination of such scores.
With the objective of maximizing the F1, we select:
\begin{equation}
\label{eq:single-selection}
\text{Single model:}\quad
d^\star \;=\; \operatorname*{arg\,max}_{m\in\{1,\dots,M\}} F1(d_m)
\end{equation}
\begin{equation}
\label{eq:ensemble-selection}
\text{Ensemble:}\quad
S^\star \;=\; \operatorname*{arg\,max}_{S\subseteq\{1,\dots,M\}} \; F1\!\left(E^{(S)}\right)
\end{equation}
Here $E^{(S)}$ denotes the stacking ensemble built from subset $S$.

\subsection{RAMSeS Overview}
\label{sec:Overview}
\begin{figure}[htbp]
    \centering
    \includegraphics[width=0.9\linewidth,
  height=0.9 \textheight,
  keepaspectratio]{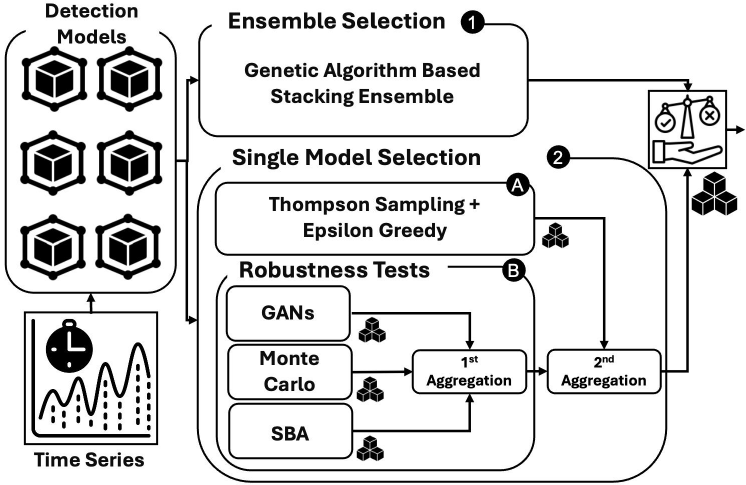}
    \vspace{-3mm} 
    \caption{RAMSeS framework overview.}
    \label{fig:Architecture}
    \vspace{-3mm}
\end{figure}

Fig.~\ref{fig:Architecture} presents \system. From a shared pool of base models, our framework proceeds along two branches: an ensemble selection and a single-model selection branch. Together, these branches provide application-specific flexibility.



\textit{The need for two-branch design.} The two branches in \system serve complementary but distinct purposes. The ensemble branch targets robustness and generalization across heterogeneous time series by leveraging model diversity, whereas the single-model branch focuses on efficiency and interpretability. Integrating both into a single unified branch would require reconciling conflicting objectives: ensemble optimization favors diversity and cross-model aggregation, while model selection prioritizes decisive ranking and parsimony. \system adaptively switches between robustness-oriented and latency-oriented configurations without retraining. 

\subsubsection{Ensemble Branch {\Large \ding{182}} in Fig.~\ref{fig:Architecture}}
\system employs a self-supervised approach that combines a meta-learner and a genetic algorithm (GA)~\cite{wolpert1992stacked, ChenLCCXZ26stacked, DamanikL25stacked, treder2024ensemble, cruz2015ensemble}. Rather than relying on simple uniform (AVG, MAX) or manual selection, the GA systematically searches for high-performing \emph{subsets of base models}. Each candidate ensemble combines the outputs of selected base models via a \emph{fixed} meta-learner, chosen \emph{a priori} based on empirical best practices. Namely, we use Random Forest (RF) as a meta-learner~\cite{Breiman01randomforest, Biau12randomforest, sun2024metarandomforest, tacsci2019metarandomforest}, in our framework, other meta-learners can be used, e.g., Logistic Regression~\cite{HosmerL00logisticregressiobook, Conklin02logisticregression, ke2023metalogisticregression, khan2020metalogisticregression}, or Support Vector Machines (SVM)~\cite{cortes1995svm, milenova2005svm, yao2023svmmetalearners, claesen2014svmmetalearners, LiaoLSL25stacked}. The meta-learner is fixed within the time series for all optimized subsets, as searching for a different meta-learner for each ensemble may increase the probability of overfitting~\cite{ChenLCCXZ26stacked, LiaoLSL25stacked}. 
The ensemble selection module outputs an optimized subset of base models.

\subsubsection{Single-Model Selection Branch {\Large \ding{183}} in Fig.~\ref{fig:Architecture}}
In parallel to the ensemble branch, \system searches for individual models that perform well on time series. 
For this purpose, it searches in parallel through linear Thompson sampling and robustness-based test (\protect\circled{A} and \protect\circled{B} in Fig.~\ref{fig:Architecture}). The outputs of both approaches are then merged into an aggregated ranking.
\paragraph{LinTS \protect\circled{A} in Fig.~\ref{fig:Architecture}}
To enable adaptive model selection under uncertainty, \system formulates the task as a contextual bandit and applies \emph{LinTS} with $\varepsilon$-greedy exploration. Each candidate model is treated as an arm. Selection is conditioned on contextual features extracted from the input stream, allowing \system to adaptively identify effective models based on localized feedback and historical performance~\cite{agrawal2013thompson}.
    
\paragraph{Robustness-based tests \protect\circled{B} in Fig.~\ref{fig:Architecture}}
Three parallel heuristics perturb time series to rank models based on distinct aspects of robustness and sensitivity. First, \emph{GAN-based perturbation testing} evaluates robustness against semantically plausible perturbations by generating \emph{borderline}, hard-to-detect anomalies across all time series~\cite{wang2017generative, goodfellow2014generative}. Second, \emph{SBA analysis} examines sensitivity near decision boundaries by injecting anomalies via local-statistics–scaled noise around threshold regions~\cite{lindemann2021survey}. Third, \emph{MC simulations} test stability under diverse anomaly patterns through MC–based brute-force tests using multiple randomized anomaly settings~\cite{SadrZH23_monte_carlo,Raychaudhuri08_monte_carlo}.
As shown in Fig.~\ref{fig:Architecture}, resulting robustness rankings are then combined through \emph{Markov-based rank aggregation} into a single ranking, presenting the first aggregation. Our Markov ranking approach is described in more detail in ~\ref{sec:markov-agg}.
Finally, we use another Markov aggregation approach, to obtain one unified ranking among both single model selection branches.

\begin{figure}[htbp]
    \centering
    \includegraphics[width=1\linewidth,
  height= \textheight,
  keepaspectratio]{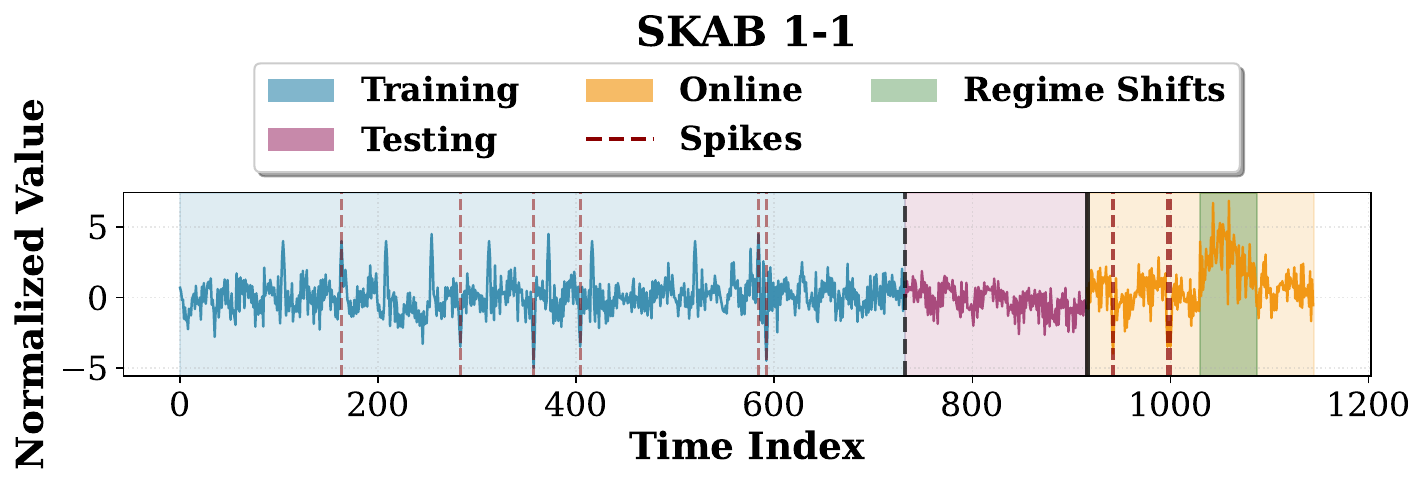}
    \vspace{-8mm} 
    \caption{SKAB 1-1 time series}
    \label{fig:skab_timeseries}
    \vspace{-3mm}
\end{figure}

\noindent\textbf{Example: SKAB 1-1 Water Circulation}
\label{subsec:end2endexample}To illustrate \system's operation, we use \texttt{SKAB 1-1}~\cite{skab}, a water circulation monitoring dataset with 8 sensor features including: accelerometers, flow rate, pressure, temperature, voltage. The time series exhibits a pump failure anomaly where flow rate drops from 32 L/min to 3.2 L/min with correlated temperature decrease. We use 8 pre-trained base models: 1 RNN, 1 MD, 1 DGHL, 1 CBLOF, 2 KNN, and 2 LOF variants. As depicted in Fig.~\ref{fig:skab_timeseries}, the dataset is split into offline (train+test) and online deployment with re-optimization every $N=5$ windows.

\section{Ensemble Selection}
\label{sec:ensemble}
\system's ensemble branch searches subsets of base models and trains a meta-learner to produce a stacked ensemble.

\begin{figure}[htbp]
    \centering
    \includegraphics[  width=0.9\linewidth,
  height=0.8 \textheight,
  keepaspectratio]{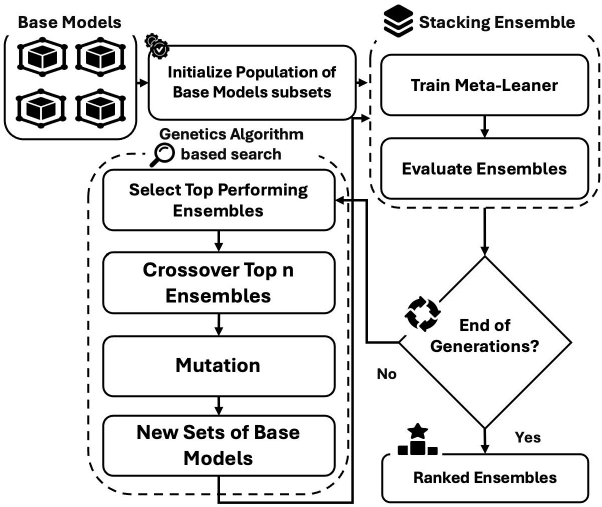}
    \vspace{-3mm} 
    \caption{Stacking ensemble via a genetic algorithm.}
    \label{fig:Ens_GA}
    \vspace{-6mm}
\end{figure}

\textit{Design rationale.}
Stacking combines different detection models with complementary strengths by training a meta-learner on their outputs~\cite{wolpert1992stacked}. A stacking ensemble comprises level-0 base learners whose predictions are used as features for a level-1 meta-learner~\cite{ChenLCCXZ26stacked}. Fixing the meta-learner \emph{a priori} restricts the search space to the possible combinations of base-model subsets, which (i) reduces overfitting risk, (ii) stabilizes the optimization, and (iii) makes convergence behavior interpretable. The ensemble search space is a discrete subset-selection problem over heterogeneous base models. 

Existing solutions for searching ensemble space use grid search or bayesian optimization. 
\emph{Grid search} scales with combinatorial complexity as $\mathcal{O}(2^M)$ or $\binom{M}{k}$ for fixed sizes and becomes infeasible even at moderate number of models $M$~\cite{AlibrahimL21_gen_vs_grid_bayes}. 
\emph{Bayesian optimization} works best for low-dimensional continuous hyperparameters with a smooth response surface. However, our ensemble selection task involves choosing discrete subsets of heterogeneous models, which violates these assumptions and requires custom kernels over sets. This causes slow convergence~\cite{AlibrahimL21_gen_vs_grid_bayes}. 
Our goal is to avoid searching all subsets and take advantage of the predictions by the meta-learner. 
Considering the possible subsets of models as successive genes, where any extension of a model subset is the child of another, a genetic algorithm focuses on traversing high-performing subset generations. This way, the number of evaluations reduces drastically and converges faster~\cite{AlibrahimL21_gen_vs_grid_bayes}. 

\begin{algorithm}[htbp]
\caption{GA based stacking-ensemble}
\label{algorithm: GA_Ensemble_Optimization_Meta_Learners}
\scriptsize
\SetAlgoLined
\SetNlSty{}{}{\,.}
\SetAlgoNlRelativeSize{-1}
\DontPrintSemicolon

\KwIn{
    $G$: \#generations,    $\mathcal{D}$: set of pre-trained base models, $P$: population size, $T$: validation data, $N$: selection number, $\mu$: mutation rate, $\phi$: fixed meta-learner
}
\KwOut{
    $\hat{S}$: optimized model subset with trained meta-learner \\
    $J_{\sigma}\!\big(E^{(\hat{S})}\big)$: best objective value (F1/AUC\mbox{-}PR)
}
\BlankLine

\lForEach{$i \in \{1, \dots, P\}$}{
    $\mathcal{S}_0[i] \leftarrow$ random subset from $\mathcal{D}$ {\hypertarget{ga:init}{\textcolor{ForestGreen}{(initialization)}}}
}

\For{$g \leftarrow 1$ \KwTo $G$}{
    \ForEach{$S \in \mathcal{S}_{g-1}$}{
        Train $\phi$ on stacked base-model outputs of $S$ over $T$  {\hypertarget{ga:training}{\textcolor{ForestGreen}{(training)}}} \;
        Compute $F1$ and/or $AUC-PR$ on the validation fold \;
    }
    $\mathcal{S}_{\mathrm{elite}} \leftarrow$ top-$k$ subsets by $J_{\sigma}$ \; 

    \For{$i \leftarrow 1$ \KwTo $P$}{
        $S_1, S_2 \leftarrow$ select $N$ parents from $\mathcal{S}_{\mathrm{elite}}$  {\hypertarget{ga:selection}{\textcolor{ForestGreen}{(selection)}}}\;
        $S' \leftarrow \text{crossover}(S_1, S_2)$ {\hypertarget{ga:crossover}{\textcolor{ForestGreen}{(crossover)}}}\; 
        \If{$\mathrm{rand}() < \mu$}{
            $S' \leftarrow \text{mutate}(S')$  {\hypertarget{ga:mutation}{\textcolor{ForestGreen}{(mutation)}}}\;
        }
        add $S'$ to $\mathcal{S}_g$ \;
    }

    $\hat{S}_g \leftarrow \operatorname*{arg\,max}_{S \in \mathcal{S}_g} J_{\sigma}\!\big(E^{(S)}\big)$ \;
}
\Return{$\hat{S}_G$, $J_{\sigma}\!\big(E^{(\hat{S}_G)}\big)$}
\end{algorithm}
\vspace{-3mm}

\subsection{Optimization procedure.}
The workflow of the GA approach is depicted in Figure~\ref{fig:Ens_GA} and formalized in Algorithm~\ref{algorithm: GA_Ensemble_Optimization_Meta_Learners}.
The GA iterates through a population of candidate subsets that are evaluated by the meta-learner in a self-supervised manner.
To start with a representative population, the GA initializes a population $\{S^{(p)}\}$ by randomly sampling distinct subsets from $\mathcal{D}$ (\hyperlink{ga:init}{initialization} in Algorithm~\ref{algorithm: GA_Ensemble_Optimization_Meta_Learners}). 
For each $S^{(p)}$, \system trains the fixed meta-learner $\phi$ on stacked base-model outputs. It then computes $F1$ and/or $AUC-PR$ for each subset on a held-out validation fold (\hyperlink{ga:training}{training} in Algorithm~\ref{algorithm: GA_Ensemble_Optimization_Meta_Learners}).
The GA then selects the top-$n$ parent ensembles based on $F1$ and/or $AUC-PR$ for the crossover step. In this step,  the models of the selected parent subsets are crossed to form new subsets (\hyperlink{ga:selection}{selection} in Algorithm~\ref{algorithm: GA_Ensemble_Optimization_Meta_Learners}).
Finally, the \textit{Mutation} step injects exploratory variations, such as adding, removing, or replacing base models, in a subset. The mutation rate is chosen based on best practices and can be adaptively adjusted based on observed convergence dynamics (\hyperlink{ga:mutation}{mutation} in Algorithm~\ref{algorithm: GA_Ensemble_Optimization_Meta_Learners}).
This process repeats for a fixed number of generations. In the end, the best-scoring subset $\hat{S}$ is the chosen ensemble.

\noindent \textbf{Example.} On SKAB 1-1~\ref{subsec:end2endexample}, the GA-Ens branch initializes a population of 20 ensembles with RF as a meta-learner and $\mu=0.1$. GA converges within 20 generations on ensemble \texttt{\{RNN, KNN\_2, LOF\_2, CBLOF\}}. This diversity captures complementary failure modes: RNN models temporal flow patterns during normal operation (32 L/min steady-state), LOF\_2 detects density shifts when flow drops abruptly, CBLOF identifies cluster deviations in the 8-sensor space, and KNN\_2 captures local proximity patterns. The meta-learner learns to weight LOF\_2 and CBLOF higher during multi-sensor anomalies (flow-temperature correlation breakdown) and RNN during stable temporal regimes.


\section{Single Model Selection}
\label{sec:single-model-selection}
The single model selection branch of \system has two components: (i) a \emph{LinTS} with $\varepsilon$ greedy module to produce model ranking, and (ii) a \emph{robustness and sensitivity} module that derives complementary rankings. Both rankings are aggregated using \emph{Markov-based rank aggregation} to obtain a consensus ordering. Next, we will explain each part in detail.

\subsection{Linear Thompson Sampling (LinTS)}
\label{sec:LinTS}

\begin{figure}[t]
    \centering
    \includegraphics[   width=0.9\linewidth,
  height=0.8 \textheight,
  keepaspectratio]{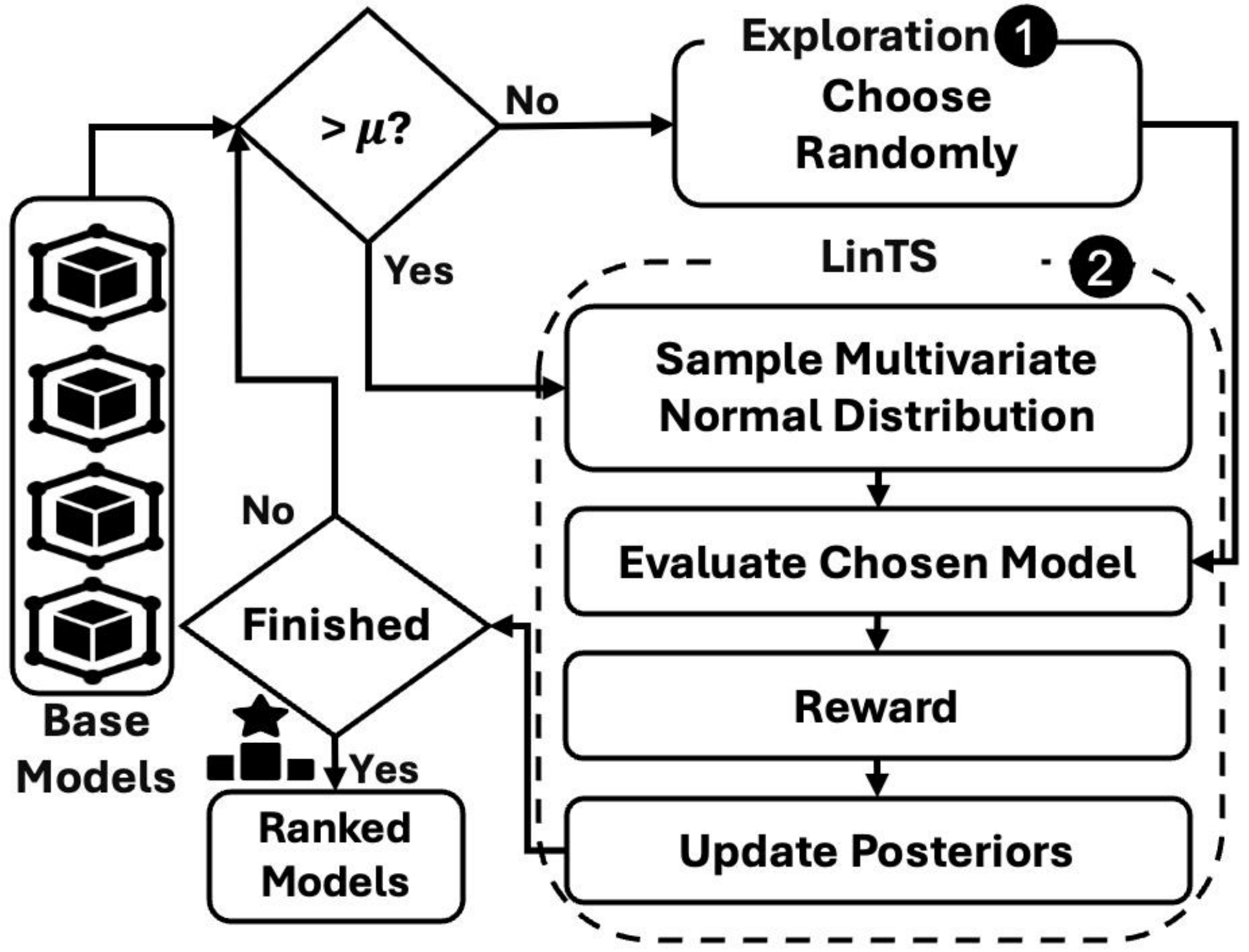}
    \vspace{-4mm} 
    \caption{Linear Thompson sampling with $\varepsilon$-greedy for model selection.}
    \label{fig:LTs_Epsilon_Architecture}
    \vspace{-3mm} 
\end{figure}

LinTS with $\varepsilon$-greedy is a lighweight and uncertainty-aware approach to balance models exploration and exploitation. In our setting, each sliding window yields immediate rewards (F1 and/or AUC-PR) that reflect local data characteristics. The Bayesian posterior over model rewards captures both performance and uncertainty, guiding the selection toward models with high expected reward and strong evidence.

\begin{algorithm}[t]
\caption{LinTS with $\varepsilon$-greedy}
\label{alg:lts_epsilon}
\scriptsize
\SetAlgoLined
\SetNlSty{}{}{\,.}
\SetAlgoNlRelativeSize{-1}
\DontPrintSemicolon

\KwIn{
    $X$: time-series stream,
    $\mathcal{D}$: set of candidate models,
    $w$: window size for segmentation,
    $\varepsilon_0$: initial exploration probability,
    $\kappa$: exploration decay rate,
    $N$: number of evaluation windows,
    $\alpha$: reward weight for F1 vs.\ AUC\mbox{-}PR
}
\KwOut{
    $\hat{m}$: models ranking based on posterior means
}
\BlankLine

\textbf{Initialize:} \\
Segment $X$ into windows $\{W_t\}_{t=1}^N$ \;
For each model $m \in \mathcal{D}$, initialize posterior parameters: \\
\hspace{1.5em} $\mu_m \leftarrow \vec{0}$,\quad $\Sigma_m \leftarrow I$ \;
Set $\varepsilon_t \leftarrow \varepsilon_0$ {\hypertarget{LinTS:init}{\textcolor{ForestGreen}{(initialization)}}}\;

\For{$t \leftarrow 1$ \KwTo $N$}{
    \textbf{Sample model $m_t$} using a hybrid strategy: \\
    \hspace{1.5em} With probability $\varepsilon_t$: select $m_t \sim \text{Uniform}(\mathcal{D})$ \\
    \hspace{1.5em} Else: sample $\tilde{\theta}_m \sim \mathcal{N}(\mu_m, \Sigma_m)$ for each $m \in \mathcal{D}$; choose $m_t = \arg\max_m \tilde{\theta}_m^\top x_t$ {\hypertarget{LinTS:selection}{\textcolor{ForestGreen}{(selection)}}}\;

    \textbf{Evaluate} $m_t$ on $W_t$ and compute reward: \\
    \hspace{1.5em} $r_t \leftarrow \alpha \cdot \mathrm{F1}(m_t, W_t) + (1-\alpha)\cdot \mathrm{AUC\mbox{-}PR}(m_t, \mathcal{B}_t)$ \;
    \hspace{1.5em} \textit{\% where $\mathcal{B}_t$ is a rolling buffer of recent windows} {\hypertarget{LinTS:reward}{\textcolor{ForestGreen}{(reward)}}}\;

    \textbf{Update posteriors} for $m_t$ via Bayesian linear regression: \\
    \hspace{1.5em} $\Sigma_{m_t}^{-1} \leftarrow \Sigma_{m_t}^{-1} + x_t x_t^\top$ \\
    \hspace{1.5em} $\mu_{m_t} \leftarrow \Sigma_{m_t} \left( \Sigma_{m_t}^{-1} \mu_{m_t} + x_t r_t \right)$ {\hypertarget{LinTS:post}{\textcolor{ForestGreen}{(Posterior Update)}}}\;

    \textbf{Anneal exploration:} \\
    \hspace{1.5em} $\varepsilon_{t+1} \leftarrow \varepsilon_0 \cdot \exp(-\kappa t)$ {\hypertarget{LinTS:anneal}{\textcolor{ForestGreen}{(annealing)}}}
}
\Return{$\hat{m}$}
\end{algorithm}

\textit{Design rationale.}
We adopt LinTS because it extends classical bandit formulations with contextual awareness through a Bayesian linear reward model~\cite{agrawal2013thompson, BanH20_bandit, DingLL19_bandit}.
The algorithm begins with sliding-window preparation and alternates between exploration, random choice with probability $\varepsilon_t$, and exploitation, posterior sampling and maximization. Each model is scored via F1 and/or AUC\mbox{-}PR rewards, and Bayesian posterior updates guide subsequent selections. The overall architecture is shown in Figure~\ref{fig:LTs_Epsilon_Architecture} and detailed in Algorithm~\ref{alg:lts_epsilon}.

\textit{Priors and initialization.}
For each model $i$, reward parameter $\theta_i\in\mathbb{R}^d$ follows a Gaussian prior
$\theta_i \sim \mathcal{N}(\mu_{i,0},\,\Sigma_{i,0})$ with mean $\mu_{i,0}=\mathbf{0}$ and covariance $\Sigma_{i,0}=\lambda^{-1}I_d$.
Initializing with a ridge prior ($\lambda>0$) stabilizes early updates and ensures $\Sigma_{i,t}$ remains positive definite. Rewards are normalized to $[0,1]$ before updating the posterior so that a unit-variance likelihood assumption holds, (\hyperlink{LinTS:init}{initialization} in Algorithm~\ref{alg:lts_epsilon}).

Our LinTS algorithm operates as follows. For each window $W_t$, we construct a contextual representation $x_t \in \mathbb{R}^d$. Each model $i$ maintains a Bayesian linear reward model
\begin{equation}
    \mathbb{E}[r_t^{(i)} \mid x_t] = \theta_i^\top x_t,
\end{equation}
with model-specific posterior parameters $(\mu_{i,t},\Sigma_{i,t})$. As mentioned in the \hyperlink{LinTS:selection}{selection step} of Algorithm~\ref{alg:lts_epsilon}, at time $t$, with probability $\varepsilon_t$ we select a model uniformly at random (exploration), otherwise, we draw $\tilde{\theta}_i \sim \mathcal{N}(\mu_{i,t},\Sigma_{i,t})$ and select $\arg\max_i \tilde{\theta}_i^\top x_t$ (exploitation via LinTS). The exploration rate decays as $\varepsilon_t = \varepsilon_0 \exp(-\kappa t)$. For the selected model $m_t$ applied to $W_t$, we compute the reward
\begin{equation}
    r_t = \alpha \cdot \mathrm{F1}(m_t, W_t) + (1-\alpha)\cdot \mathrm{AUC\mbox{-}PR}(m_t, W_t),
\end{equation}
where $\alpha \in [0,1]$ weights thresholded performance (F1) versus threshold-free ranking (AUC\mbox{-}PR). As depicted in \hyperlink{LinTS:reward}{reward step} in Algorithm~\ref{alg:lts_epsilon}, LinTS computes rewards \emph{exclusively} on batches of injected windows with synthetic labels $\tilde{y}$. After observing $(x_t,r_t)$, we update the posterior via
\begin{align}
    \Sigma_{i,t+1} &= \big(\Sigma_{i,t}^{-1} + x_t x_t^\top\big)^{-1} \label{eq:posterior_covariance} \\
    \mu_{i,t+1} &= \Sigma_{i,t+1}\big(\Sigma_{i,t}^{-1}\mu_{i,t} + x_t r_t\big). \label{eq:posterior_mean}
\end{align}
After processing $N$ windows, models are ranked according to their posterior rewards, and this LinTS ranking is passed to the final aggregation along with the robustness-based ranking.

\textit{Cold start.}
At initialization, higher $\varepsilon_0$ encourages broad exploration. $\varepsilon_t$ then decreases to prioritize exploitation. To stabilize rewards in short windows, we optionally apply exponential smoothing to $r_t$ before updating the posterior. Exponential smoothing averages recent rewards with past ones, giving higher weight to recent observations to reduce noise.

\noindent \textbf{Example.} Applying LinTS to SKAB 1-1, it explores all 8 models over 50 validation windows and selects the \texttt{RNN} based on accumulated posterior means. This selection aligns with the properties of SKAB 1-1 from Figure~\ref{fig:skab_timeseries}: \texttt{RNN}'s sequential architecture captures temporal dependencies in pump flow, pressure, and temperature dynamics, outperforming proximity-based methods, such as \texttt{LOF} or \texttt{KNN}, that ignore temporal correlations between the 8 sensors.

\subsection{Robustness and Sensitivity Tests}
\label{subsec:robustness}
Models robustness and sensitivity is evaluated via three complementary tests. Each test yields a model ranking which is merged by Markov-based rank aggregation.
\begin{algorithm}[t]
\caption{GAN-based perturbations}
\label{alg:gan_training}
\small
\SetAlgoLined
\SetNlSty{}{}{\,.}
\SetAlgoNlRelativeSize{-1}
\DontPrintSemicolon

\KwIn{$data$: original data; $labels$: original labels; $input\_dim$: input dimensions; $epochs$; $batch\_size$: batch size; $noise\_dim$: noise dimensions}
\KwOut{updated data, labels, indices of injected points}

\BlankLine
\textbf{Initialize} generator and discriminator \;
\textbf{Train} with label smoothing and noise for $epochs$ \;
\textbf{Generate} borderline points with the trained generator \;
\textbf{Score} candidates with the discriminator and assign labels \;
\textbf{Integrate} generated points via sliding windows \;
\textbf{Return} updated arrays and injection indices \;
\end{algorithm}

\subsubsection{GAN-Based Robustness Evaluation}
This module tests models against anomalies that mimic realistic drifts while preserving temporal structure.

\textit{Architecture and training.}
We instantiate a generator $G$ and discriminator $D$ as two-layer Multi-Layer Perceptrons (MLPs)~\cite{goodfellow2014generative, wang2017generative} with 256 hidden units, ReLU activations, and dropout for regularization.
The generator maps $G:\mathbb{R}^{d}\!\rightarrow\!\mathbb{R}^{d}$ with a \texttt{tanh} output layer, while the discriminator maps $D:\mathbb{R}^{d}\!\rightarrow\![0,1]$. Optimization follows Algorithm~\ref{alg:gan_training} using binary cross-entropy losses and Adam optimizers with a learning rate of $10^{-4}$. To improve training stability, we apply label smoothing for real and fake targets and add Gaussian noise to both. Training proceeds for a fixed number of epochs with mini-batches, and seeds are fixed to ensure reproducibility.

\textit{Data preparation.}
GAN training operates on the same per-window data stream. To match the generator’s $\tanh$ output layer, inputs are linearly scaled to $[-1,1]$. GAN training uses the clean, non-augmented, split to avoid leakage. Augmentation occurs \emph{after} training, during robustness testing.

\textit{Injection.}
After training, we generate a candidate pool of points $\mathcal{C}=\{x^{(g)}_k\}_{k=1}^K$ with $x^{(g)}_k=G(z_k)$, $z_k\sim\mathcal{N}(0,I)$. We score candidates with $D$ and measure ambiguity against the decision boundary via
\begin{equation}
\label{eq:gan_ambiguity}
\delta_k \;=\; \bigl\lvert D\!\bigl(x^{(g)}_k\bigr) - \tau \bigr\rvert
\end{equation}
where $\tau$ denotes the discriminator decision threshold separating normal from anomalous samples.
We then select the $B$ most ambiguous points by minimizing the total ambiguity:
\begin{equation}
\label{eq:gan_topB}
\mathcal{I}_B^\star \;=\; \operatorname*{arg\,min}_{\mathcal{I}\subseteq\{1,\dots,K\},\,|\mathcal{I}|=B}\; \sum_{k\in\mathcal{I}} \delta_k,
\qquad
\mathcal{X}_B^\star \;=\; \{x^{(g)}_k : k\in\mathcal{I}_B^\star\}.
\end{equation}
For surrogate labels we use the same boundary:
\begin{equation}
\label{eq:gan_labels}
\hat{y}(x) \;=\; \mathbb{I}\!\bigl[ D(x) \ge \tau \bigr], \qquad x\in\mathcal{X}_B^\star,
\end{equation}
yielding an augmented set that contains both near-normal ($\hat{y}{=}0$) and near-anomalous ($\hat{y}{=}1$) behaviors.

\textit{Temporal integration policy.}
To respect chronology, we interleave the selected borderline points $\mathcal{X}_B^\star$ into the stream at regular intervals within sliding windows, using an injection budget $\rho$, default $\rho\approx 0.1$ of the original number of samples. Integration preserves ordering, updates label masks, and records injection indices for traceability and visualization.

\textit{Evaluation and ranking.}
Each candidate model is re-evaluated on the GAN-augmented data. Then we compute F1 and/or AUC\mbox{-}PR and produce model ranking. This ranking forms one of the outputs of the robustness block.

\subsubsection{Off-by-Threshold (SBA)}
This analysis measures sensitivity to deviations from normality by injecting statistical border points, based on local context.
Given a time series $X \in \mathbb{R}^{d\times n}$, SBA injects synthetic anomalies at regular intervals by adding zero-mean Gaussian noise with per-feature scaling derived from a contextual window of length $w_{\text{ctx}}$. For each injected point, we draw
\begin{equation}
\label{eq:offby_noise}
\tilde{x}_j \sim \mathcal{N}\!\bigl(0,\,\sigma_j^2 s^2\bigr), \quad s \sim \mathcal{U}[\gamma_{\min},\gamma_{\max}],
\end{equation}
where $\sigma_j$ is the local standard deviation for feature $j$ in the contextual window. Points with $s \le 1$ are labeled normal ($y{=}0$), while points with $s>1$ are labeled as statistical border anomalies ($y{=}1$), creating a dataset with known labels near decision boundaries. We then evaluate all candidate models on this augmented data, compute F1 and AUC\mbox{-}PR, and derive an SBA ranking that identifies models robust to near-threshold variations. The perturbed data is not re-used in other components of our system.

\subsubsection{Monte Carlo Simulation (MC)}
We perform Monte Carlo stress testing to evaluate model stability under diverse anomaly patterns. In each of $R$ randomized trials, synthetic anomalies of varying magnitudes ($s \in [\gamma_{\min}, \gamma_{\max}]$) are injected at random positions in the input streams, and all models are re-evaluated. The resulting F1 and/or AUC-PR scores are averaged across trials to obtain a robust Monte Carlo ranking~\cite{raychaudhuri2008introduction}.

\subsection{Markov-Based Rank Aggregation}
\label{sec:markov-agg}
We aggregate multiple rankings over the same model set $\mathcal{D}=\{1,\dots,D\}$ into a single consensus ordering via a Markov-chain construction from pairwise preferences~\cite{DuPJDPYX22_markov}. Given rankings $\mathcal{R}=\{r^{(1)},\dots,r^{(K)}\}$, we count preferences to form $C\in\mathbb{R}^{D\times D}$, where $C_{ij}$ is the number of times model $i$ is ranked ahead of $j$. We then obtain a row-stochastic transition matrix $P$ by normalizing rows:
\begin{equation}
\label{eq:markov-P}
P_{ij} \;=\;
\begin{cases}
\dfrac{C_{ij}}{\sum_{\ell\neq i} C_{i\ell}}, & \text{if } j\neq i \text{ and } \sum_{\ell\neq i} C_{i\ell}>0,\\[0.9ex]
0, & \text{if } j=i \text{ and } \sum_{\ell\neq i} C_{i\ell}>0,\\[0.6ex]
\dfrac{1}{D}, & \text{if } \sum_{\ell\neq i} C_{i\ell}=0,
\end{cases}
\end{equation}
The consensus score is the stationary distribution $v^\star$ satisfying $v^\star=v^\star P$, which we compute by power iteration from a uniform initialization until $\ell_1$ convergence. Models are finally sorted in \emph{descending} order of $v^\star$ to produce the consensus ranking. We apply this operation (i) to aggregate the three robustness rankings, GAN, off-by-threshold, Monte Carlo, and (ii) to aggregate the robustness with the LinTS ranking.

\subsection{Robustness Evaluation Mechanism}
\label{subsec:robustness_mechanism}
We evaluate the robustness of base models via three tests that probe orthogonal modes:
(I) As explained in Equation~\ref{eq:gan_ambiguity}, the \textit{GAN} learns from clean data and produces points near the discriminator's decision boundary. Models that only memorize training patterns fail on these cases. (II) \textit{SBA} tests sensitivity of base models by injecting Gaussian noise scaled by local standard deviation within overlapping windows. Unlike GAN's distribution-learned perturbations, SBA uses context-dependent noise to probe decision boundaries under measurement uncertainty, for example peak-period temperature fluctuations near the threshold. (III) The \textit{Monte Carlo simulation} tests stochastic stability through random perturbations. Providing a randomized baseline independent of GAN or SBA approaches.

\textit{Preventing circularity.} Models are never trained on components perturbed augmented data. As depicted in Figure~\ref{fig:Architecture}, each method operates on a data copy, and we aggregate model rankings via Markov aggregation (Equation~\ref{eq:markov-P}), which combines rankings from all three tests into a consensus ranking.

\noindent \textbf{Example.} On SKAB 1-1, we observe the following behaviour: With 100 epochs the \textit{GAN} generates perturbed flow-temperature pairs near the 32 L/min operating point, choosing \texttt{LOF\_2} for its resilience to borderline cases. \textit{SBA}: injects near-threshold degradation (31-33 L/min), also selecting \texttt{LOF\_2} for boundary sensitivity. \textit{MC} adds noise across all features, e.g., accelerometer vibration during pump failure, selecting \texttt{CBLOF}.
After Markov aggregation, \texttt{LOF\_2} is selected as the single-model candidate. This divergence from LinTS's choice (\texttt{RNN}) is instructive: while RNN excels on normal temporal patterns (32 L/min steady-state), it fails to generalize to regimes absent from training. LOF\_2's density-based approach treats this as a sparse region in the 8-dimensional sensor space, detecting multi-sensor correlation shifts (flow-temperature-vibration coupling) that violate normal operational density.

\subsection{Online Anomaly Detection and Adaptive Model Selection}
\label{subsection:online}
\begin{figure}[t]
\centering
\includegraphics[  width=0.9\linewidth,
  height=0.9 \textheight,
  keepaspectratio]{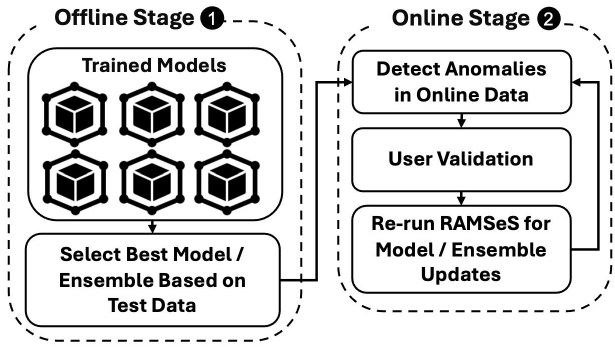}
  \vspace{-4mm} 
\caption{Continuous online real-time model selection for anomaly detection.}
\label{fig:realtime_model_selection}
\vspace{-6mm}
\end{figure}

In \system, online real-time anomaly detection operates in a continuously adaptive loop. At initialization, both the top-ranked single model and the optimized ensemble (learned offline) are deployed.
For each new batch of incoming data, both candidates emit anomaly decisions. These can be presented for human/system feedback. Figure~\ref{fig:realtime_model_selection} shows the complete cycle, initialization, detection, feedback, and refinement. 

\textit{Window-based adaptation mechanism.} 
To implement the online adaptive loop, the time series is partitioned into 80\% for offline model selection and 20\% for online inference. The online portion is further divided into overlapping windows of size $w = 0.05 \times \text{online\_size}$ with step size $s = 0.05w$ (95\% overlap) to preserve temporal context. Every $N$ windows, the system triggers re-optimization: a sliding window is constructed by concatenating the $N$ most recent online windows while dropping an equal number of samples from the beginning of the offline data, maintaining constant training size and oracle knowledge~\cite{sylligardos2023choose, boniol2021sand}. This sliding mechanism adapts to distribution shifts while preventing catastrophic forgetting of earlier patterns. During This re-optimization step the selected ensemble or model might be updated based on performance.

\noindent\textbf{Example.} During the online phase on SKAB 1-1, \system deploys the ensemble \texttt{\{RNN, KNN\_2, LOF\_2, CBLOF\}} and the single model \texttt{LOF\_2}. At timestamp 1030, the pump failure anomaly occurs: Volume Flow drops from 32 L/min to 3.2 L/min coupled with Temperature decrease from 75°C to 74°C. The ensemble branch detects this multi-sensor correlation with F1=0.87 by combining LOF\_2's density detection with CBLOF's cluster analysis. The single-branch LOF\_2 fails to detect the anomaly (F1=0.52), missing the subtle temperature correlation. Despite LOF\_2's theoretical advantage for density shifts, the isolated single-model deployment lacks the multi-sensor context that the ensemble provides through CBLOF's cluster analysis. After $N=5$ windows, \system triggers re-optimization: the ensemble branch re-runs GA on the sliding window (concatenating recent online data while dropping early offline samples), confirming \texttt{\{RNN, KNN\_2, LOF\_2, CBLOF\}} remains optimal. The single-branch re-executes LinTS and robustness tests, switching to \texttt{CBLOF} as the top choice based on its superior performance on cluster deviations observed in recent windows.

\section{Experiments and Results}
\label{sec:results}
We analyze the strengths and weaknesses of \system with experiments on several well-known benchmarks.

\noindent\textbf{Datasets}
We evaluate the performance of our framework on three diverse time-series datasets that vary in dimensionality, domain, and anomaly characteristics. 
The \textit{UCR Anomaly Archive (UCR)}~\cite{ucr} contains average $\sim$73k time-points from 250 univariate time series (i\_UCR) entities spanning domains, such as medicine and environment. The \textit{Server Machine Dataset (SMD)}~\cite{smd} comprises average $\sim$27k time-points from 28 server machines (SMD~i-j) in a large-scale internet services infrastructure (38-dimensions). Each machine forms an entity with 38 features capturing CPU usage, memory utilization, and network throughput. The \textit{Skoltech Anomaly Benchmark (SKAB)}~\cite{skab} contains average $\sim$2k time-points, 34 time series (SKAB~i-j) sequences from water-pump operations, each with 8 sensor features recorded in normal and faulty states.

\noindent\textbf{Anomaly Detection algorithms}
\label{subsec:base-models}
Our pool of models follows the TSB-AutoAD~\cite{liu2025tsb} taxonomy in Table~\ref{tab:tsb-autoad-base-algos}, covering three families, Neural Networks, Statistical, and Foundation Models, with applicability to both univariate, and multivariate settings.
FMs are excluded from the \system candidate pool, as they showed inconsistent performance. For completeness and comparability, we include FMs only in the \textit{TSB-AutoAD} setting, which evaluates FMs in its testbed models~\cite{liu2025tsb}.

\begin{table}[t]
\centering
\caption{Base models grouped by family: Neural Networks (NN), Statistical (Stat) or Foundation Models (FM) and applicability (U/M).}
\label{tab:tsb-autoad-base-algos}

\scriptsize
\setlength{\tabcolsep}{4pt}
\renewcommand{\arraystretch}{1.05}

\begin{tabularx}{\linewidth}{@{}l c c X@{}}
\toprule
\textbf{Base models} & \textbf{Category} & \textbf{Dim} & \textbf{Brief Description} \\
\midrule
AutoEncoder \cite{SakuradaY14_AutoEncoder}          & NN   & U\&M & Reconstruction error \\
RNN \cite{ChangZHYGTCWHH17_RNN}                    & NN   & U\&M & Vanilla recurrent prediction error \\
DGHL \cite{challu2022deep_DGHL}     & NN   & M    & Gen-ConvNet with latent factor \\
LSTMVAE \cite{ParkHK18_LSTMVAE}               & NN   & U\&M & LSTM-based VAE \\
Donut \cite{XuCZLBLLZPFCWQ18_Donut}                & NN   & U\&M & VAE likelihood scoring \\
OmniAnomaly \cite{SuZNLSP19_OmniAnomaly}          & NN   & U\&M & Stochastic RNN likelihood \\
USAD \cite{AudibertMGMZ20_USAD}                 & NN   & U\&M & Adversarial autoencoders \\
TranAD \cite{TuliCJ22_TranAD}               & NN   & U\&M & Self-conditioned + adversarial \\
TimeNet \cite{WuHLZ0L23_TimeNet}              & NN   & U\&M & Temporal-variation features \\
FITS \cite{XuZ024_FITS}                 & NN   & U\&M & Frequency interpolation score \\
\midrule
ABOD \cite{KriegelSZ08_ABOD}                  & Stat & M    & Angle-based outlier factor \\
(Sub)-MCD \cite{SakuradaY14_MCD}            & Stat & U\&M & Robust Mahalanobis \\
Sub-OCSVM \cite{ScholkopfWSSP99_OCSVM}            & Stat & U\&M & One-class SVM boundary \\
RM                    & Stat & U\&M & Simple moving-average residuals\\
(Sub)-LOF \cite{breunig2000lof}            & Stat & U\&M & Local density deviation \\
(Sub)-KNN \cite{RamaswamyRS00_NN}            & Stat & U\&M & $k$NN distance score \\
KMeansAD \cite{yairi2001fault_KMeanseAD}             & Stat & U\&M & Distance to centroid \\
CBLOF \cite{HeXD03_CBLOF}                & Stat & U\&M    & Cluster-weighted LOF \\
POLY \cite{LiMM07_POLY}                 & Stat & U    & Local polynomial residuals \\
(Sub)-IForest \cite{LiuTZ08_IForest}        & Stat & U\&M & Isolation via partitions \\
(Sub)-HBOS \cite{goldstein2012histogram_HBOS}           & Stat & U\&M & Histogram bin height \\
(Sub)-PCA \cite{aggarwal2016introduction_PCA}            & Stat & U\&M & Distance to PC subspace \\
\midrule
OFA \cite{ZhouNW0023_OFA}                  & FM   & U\&M & Fine-tuned GPT-2 \\
Lag-Llama \cite{rasul2023lag_Lag-Llama}            & FM   & U    & Decoder-only with lags \\
Chronos \cite{AnsariSTZMSSRPK24_Chronos}              & FM   & U    & Pretrained T5 on TS \\
TimesFM \cite{DasKSZ24_TimeFM}              & FM   & U    & Decoder-only with patching \\
MOMENT \cite{GoswamiSCCLD24_MOMENT}               & FM   & U    & Pretrained T5 encoder \\
\bottomrule
\end{tabularx}

\vspace{0.25em}
\emph{\scriptsize Dim: U = univariate, M = multivariate. “(Sub)-” denotes a subsequence variant.}
\vspace{-5mm} 
\end{table}


\subsection{Hyperparameter Configuration}
\label{subsec:hyperparameters}
Table~\ref{tab:hyperparameters} lists all hyperparameters used in \system. To avoid overfitting to specific datasets, we tuned these hyperparameters on a randomly selected 20\% subset of time series from UCR, SMD, and SKAB. This provides sufficient diversity to capture varying characteristics, i.e., length, dimensionality, anomaly patterns. Unless otherwise noted, the selected values remain fixed across all remaining time series in our experiments.

\textit{GA-Ens.} GA with  $P\leq20$ and $G\leq20$ achieves F1 saturation while minimizing computational overhead. Mutation rate $\mu\leq0.2$ sustains diversity without degrading performance. Random Forest (RF) is chosen as the meta-learner because it achieves comparable F1 to SVM while being faster. These choices reflect fundamental GA trade-offs: $P$ controls exploration breadth, $G$ control convergence depth, and mutation balances diversity versus stability. 

\textit{LinTS.} The single-model branch uses LinTS to balance exploration and exploitation. LinTS begins with $\epsilon_0\!=\!0.2$ (20\% random exploration) and decays by $0.99$ per window, shifting toward exploitation as evidence accumulates. 
The reward function weights F1 with $\alpha\!=\!0.7$, which has been learned during parameter tuning.
A $50$-window exploration horizon provides sufficient statistical evidence for reward estimation while maintaining responsiveness to regime changes. Lower settings, e.g., $10$ windows, led to premature convergence on suboptimal models, while larger horizons, e.g., $200$ windows, delayed adaptation when anomaly patterns shifted. This aligns with guidance for stable arm selection in bandit algorithms~\cite{agrawal2013thompson}.

\textit{GAN.} Adversarial perturbations test robustness by generating borderline anomalies that follow time-series properties. The generator uses a dense layer with 256 units and dropout of 0.4 to prevent memorization of training data. Following the recommendations of ~\cite{wang2017generative, goodfellow2014generative}, both generator and discriminator are trained with Adam optimizer at learning rate $10^{-4}$. Training runs for 100 epochs to ensure stable generation of boundary cases~\cite{goodfellow2014generative}. Discriminator loss plateaus after $\sim$80 epochs for all datasets. The additional 20 epochs provide robustness and improve boundary case quality.

\textit{SBA and Monte Carlo hyperparameters.} SBA sensitivity analysis uses 10\% near-threshold augmentation to probe decision boundaries under measurement uncertainty. Monte Carlo noise injection runs $\leq$10 trials as a stochastic stress test.

\textit{Online phase hyperparameters.} Window size is set to 5\% of the online data length, ensuring adaptation to varying time series lengths. Consecutive windows overlap by 95\% using a step size of 5\% of the window size, creating a sliding window effect. Re-optimization is triggered every $\geq$5 windows to balance model freshness against computational overhead.

\begin{table}[t]
\centering

\caption{\system Hyperparameters for Reproducibility}
\label{tab:hyperparameters}
\scriptsize
\begin{tabular}{@{}llll@{}}
\toprule
\textbf{Entity} & \textbf{Parameter} & \textbf{Value} & \textbf{Rationale} \\
\midrule
\multirow{4}{*}{\makecell[l]{GA-Ens}} 
  & Population size ($P$) & $\leq$20 & F1 saturation (\ref{sec:runtime}) \\
  & Number of Generations ($G$) & $\leq$20 & Cost-effective (\ref{sec:runtime}) \\
  & Mutation rate ($\mu$) & $\leq$0.2 & Sustains diversity \\
  & Meta-learner & RF & Nonlinear aggregation \\
\midrule
\multirow{4}{*}{\makecell[l]{LinTS}} 
  & Windows & 50 & Posterior convergence \\
  & Epsilon ($\epsilon_0$) & 0.2 & Early exploration \\
  & ($\epsilon_0$) decay & 0.99 & Exploitation preferred\\
  & Reward ($\alpha$) & 0.7 & Detection accuracy \\
\midrule
\multirow{3}{*}{\makecell[l]{GAN}} 
  & Epochs & 100 & Adversarial converge \\
  & Learning rate & $10^{-4}$ & Stable training \\
  & Dropout & 0.4 & Prevent overfitting \\
\midrule
\multirow{2}{*}{\makecell[l]{SBA}} 
  & Injection factor & 0.1 & 10\% test data \\
  & Scale factor & [0.95, 1.05] & Near-threshold \\
\midrule
\multirow{2}{*}{\makecell[l]{MC}} 
  & Simulations & $\leq$10 & Statistical confidence \\
  & Noise level ($\sigma$) & 0.1 & Gaussian robustness \\
\midrule
\multirow{3}{*}{\makecell[l]{Online}} 
  & Window size & 5\% & of online data \\
  & Window step & 5\% & max overlap \\
  & Re-opt ($N$) & $\geq$ 5 windows & Adaptive balance \\
\bottomrule
\end{tabular}
\vspace{-5mm}
\end{table}

\subsection{Baselines for Comparison}
We compare \system against the top Automated TSAD architectures. 
Unless otherwise noted, we report \emph{F1} computed on test splits using the range-based event evaluation, so that sequence-level detections are assessed consistently across time series. For each time series, the ensemble and single-model selections are produced \emph{without access to labels}. 




\textit{(1) Automated Ensemble and single model selection (TSB-AutoAD~\cite{liu2025tsb}).}
We use the TSB-AutoAD framework of Liu et al.~\cite{liu2025tsb}, which provides a unified benchmark and reference implementations of automated TSAD solutions spanning model selection and model ensembling. 

\textit{(2) Single model selection (UMS)~\cite{goswami2022unsupervised}.}
UMS selects a single model by aggregating surrogate criteria prediction error, model centrality, and performance under synthetic anomaly injection.

\textit{(3) Ensemble selection (AutoTSAD~\cite{autotsad_thorsten_24}).
AutoTSAD is an unsupervised system for univariate time series that combines data generation, model tuning, and selective score ensembling.} 

\textit{(4) Meta-learning model selection (MSAD~\cite{sylligardos2023choose}).}
MSAD uses time series classification to predict which anomaly detector will perform best. A pre-trained classifier operates on extracted features from windowed segments to select a detector.

\noindent\textit{Tuning policy.} Unless otherwise stated, we do \emph{not} tune base models in \system: each base model is instantiated with a set of randomly sampled configurations, and this fixed pool is used for all experiments. 

\subsection{System Level Comparison}
\label{subsec:system_level_comparison}

Table~\ref{tab:system_level_comparison} compares \system against four TSAD baselines: TSB-AutoAD, UMS, MSAD, and AutoTSAD. All experiments were conducted on a single machine (AMD EPYC 7543P 32-Core @ 2.80 GHz, 503 GB RAM, Linux 6.1.0), using the hyperparameter values in Table~\ref{tab:hyperparameters}.

\textit{Detection Accuracy.} Table~\ref{tab:system_level_comparison} shows that \system achieves the highest F1 scores on UCR and SMD datasets, outperforming all baselines. Notably, \system demonstrates strong performance on large-scale univariate data, e.g., UCR, and high-dimensional multivariate scenarios, e.g., SMD. On SKAB, \system and UMS  achieve the highest F1 performance. SKAB's short sequences with simple patterns are well-suited to UMS's lightweight heuristics. This suggests, that while \system excels on complex, large-scale workloads, simpler methods may suffice for short, well-structured time series.

\textit{Computational Efficiency.} As depicted in Table~\ref{tab:system_level_comparison}, \system achieves faster runtime than TSB and UMS on UCR and SMD. Moreover, \system maintains moderate memory consumption (500–800 MB) across all datasets, consuming less than UMS and less than TSB on SMD and SKAB. MSAD achieves the lowest runtime on SMD and SKAB by using a pre-trained time series classifier for lightweight prediction-based selection. However, on longer time series as in UCR, MSAD is significantly slower than \system due to training latency. AutoTSAD times out on UCR after 48 hours per entity due to its exhaustive hyperparameter search for base models.

\textit{Practical Implications.} \system manages accuracy-speed-memory trade-offs across workloads, excelling on large-scale and high-dimensional scenarios. For simple datasets ($<$2k timesteps), lightweight selectors may suffice, but \system's two-branch architecture ensures robust performance when dataset characteristics, e.g., length, dimensionality, pattern complexity, are unknown a priori.

\begin{table}[t]
\centering
\caption{Average System-level performance: \system vs. automated TSAD baselines across all datasets and Time series (TS). Time (T) in seconds, Memory (M) in MB. Best in \textbf{bold}. Timeout is 48 hours per time series.}
\label{tab:system_level_comparison}
\scriptsize
\setlength{\tabcolsep}{4pt}
\begin{tabular}{l|rrr|rrr|rrr|}

\toprule
\multirow{2}{*}{\textbf{Method}} & 
\multicolumn{3}{c|}{\textbf{UCR (250 TS)}} & 
\multicolumn{3}{c|}{\textbf{SMD (27 TS)}} & 
\multicolumn{3}{c|}{\textbf{SKAB (18 TS)}} \\
\cmidrule(lr){2-4} \cmidrule(lr){5-7} \cmidrule(lr){8-10}
& \textbf{F1} & \textbf{T}  & \textbf{M}  & 
\textbf{F1}  & \textbf{T}  & \textbf{M}  & 
\textbf{F1}  & \textbf{T}  & \textbf{M}\\
\midrule
    TSB    & 0.45 & 5012 & \textbf{534} & 0.55 & 4603 & 825 & 0.68 & 368 & 702 \\
UMS           & 0.66 & 2661 & 2580 & 0.80 & 1520 & 1796 & \textbf{0.86} & 546 & 1312 \\
MSAD           & 0.07 & 14918 & 639 & 0.13 & \textbf{260} & \textbf{608} & 0.54 & \textbf{74} & \textbf{286} \\
AutoTSAD      & \multicolumn{3}{c|}{Timeout} & \multicolumn{3}{c|}{Not applicable} &  \multicolumn{3}{c|}{Not applicable}   \\
\midrule
\system & \textbf{0.83} & \textbf{1704} & 785 & \textbf{0.89} & 586 & 618 & \textbf{0.86} & 647 & 512 \\
\bottomrule
\end{tabular}%
\vspace{-3mm}
\end{table}


\subsection{Component-level Runtime and Memory Analysis}
\label{subsec:computational_overhead}
To assess the viability of \system, we report per-component runtime and system-level memory footprint across the three datasets for our offline phase.
Table~\ref{tab:computational_overhead} reports average runtime and memory across benchmark datasets. Runtime is reported by module. Markov aggregation contributes negligible overhead ($<$0.01s) and is omitted. \system's runtime and memory depend on the time series characteristics: sequence length drives the number of LinTS windows as more windows require more posterior updates, dimensionality scales meta-learner training cost in GA-Ens as higher-dimensional feature spaces increase meta-learner complexity and latency. The memory footprint is lower than 800MB for all datasets.

\begin{table}[t]
\centering
\caption{Average end-to-end runtime (seconds) and memory (MB) per entity.}
\label{tab:computational_overhead}
\scriptsize
\setlength{\tabcolsep}{3pt}
\renewcommand{\arraystretch}{1.1}
\begin{tabular}{@{}lrrrrrrr@{}}
\toprule
\textbf{Dataset} & \textbf{GA} & \textbf{LinTS} & \textbf{GAN} & \textbf{SBA} & \textbf{MC} & \textbf{Single Branch} & \textbf{System Memory} \\
\midrule
UCR    &  75  & 1445 & 71 & 39 & 73 & 1628 & 785   \\
SMD    &  280 & 173  & 38 & 30 & 64 & 306  & 618   \\
SKAB   &  134 & 337  & 75 & 42 & 88 & 513  & 512  \\
\bottomrule
\end{tabular}
\vspace{-5mm}
\end{table}

\textit{Bottleneck on UCR.} Table~\ref{tab:computational_overhead} shows that LinTS consumes 1,445 seconds on UCR, representing 84.8\% of the total single-branch runtime. This overhead stems from posterior updates over 50 windows on long univariate series with an average of 73,000 timesteps. Each window triggers a Gaussian update involving mean and covariance inversion for each candidate model. With UCR's longer sequences, the number of windows grows linearly with series length, and each covariance inversion scales cubical in the  feature dimension. Consistent with LinTS complexity~\cite{agrawal2013thompson}, posterior updates dominate runtime when the number of exploration rounds is large.

\textit{Bottleneck on SMD.} The comparably high dimensionality  of $38$ features dominates the computational cost of both base models inference and meta-learner fitting.
Table~\ref{tab:computational_overhead} shows that GA-Ens accounts for $47.7\%$ of the total runtime on SMD due to the meta-learner, which trains over $38$ features across $P\!=\!20$ models for $G\!=\!20$ generations. Each generation requires training the RF on stacked predictions from base models. 

\textit{Bottleneck on SKAB.} Table~\ref{tab:computational_overhead} shows that \system exhibits more evenly distributed costs across components on SKAB compared to UCR and SMD. The shorter sequence length, average 2,000 timesteps, reduces the number of LinTS windows, mitigating the posterior update bottleneck observed on UCR. The moderate dimensionality of 8 features, compared to 38 in SMD, reduces meta-learner training cost in GA-Ens.


\subsection{Online Inference, Scalability and Window Size Effect}
\label{subsec:online_scalability}

We evaluate \system with regard to: (1) online per-window performance, latency and memory, (2) scalability vs. candidate pool size, and (3) sensitivity to window size. All experiments use hyperparameters from Table~\ref{tab:hyperparameters}.

\subsubsection{Online Phase Performance}
\label{subsubsection:online-inference}
\begin{figure}[ht]
\centering
\includegraphics[width=0.50\textwidth]{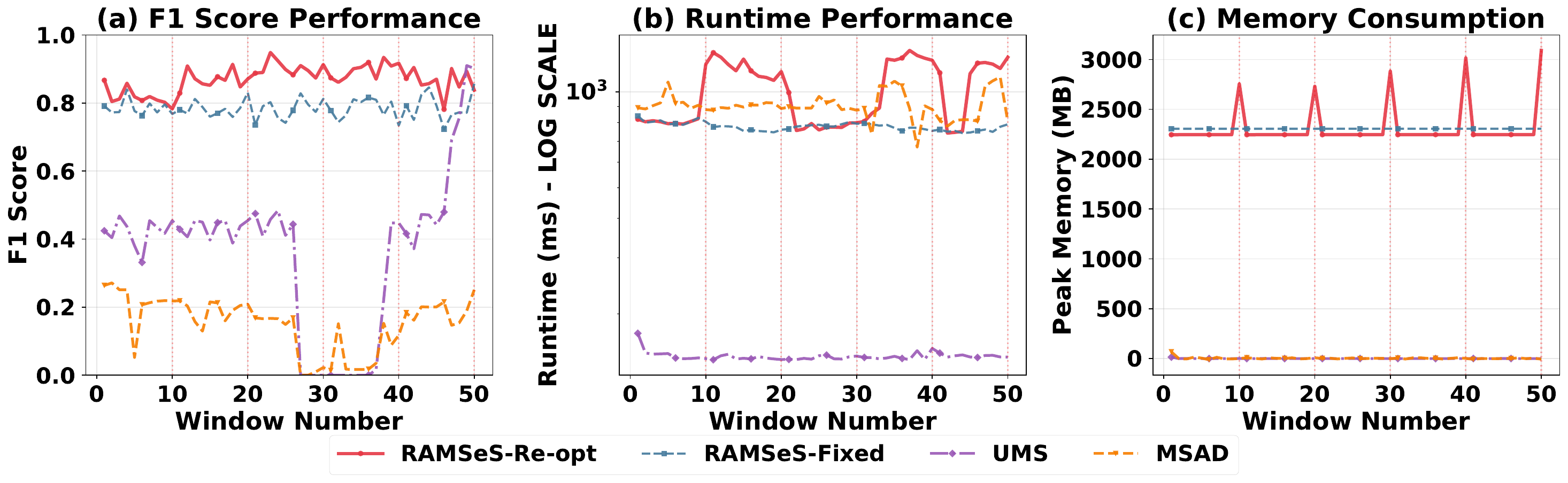}
\vspace{-7mm} 
\caption{Online phase avg. F1 \& overhead on SMD for RAMSeS \& Baselines}
\label{fig:online_phase_performance}
\vspace{-2mm}
\end{figure}

To simulate the real-time scenarios we inject 8 random regime shifts to the online stream and test the performance of \system against UMS and MSAD baseline. AutoTSAD is only applicable to univariate time-series and TSB is also excluded due to low performance, F1 $\leq$10\%.
Fig.~\ref{fig:online_phase_performance} compares average F1 score, latency, and memory consumption across 50 overlapping windows with 5\% step size, on SMD for: \system with re-optimization (every 10 windows), \system without re-optimization (fixed selection), and the baselines. Both \system variants achieve on average 40\% higher F1 than UMS and 60\% higher F1 compared to MSAD. The re-optimization variant shows average performance gain after each update cycle, most notably after window 10 (the first update), where F1 jumps relative to the fixed variant, establishing it as the highest performer. However, re-optimization incurs higher memory during updates. This trade-off suggests practitioners should enable or disable re-optimization depending on whether their use case prioritizes accuracy or resource constraints.

\subsubsection{Online Scalability vs. Candidate Pool Size}
\label{subsubsection:scalability-pool-size}
\begin{figure}[ht]
\centering
\includegraphics[width=0.50\textwidth]{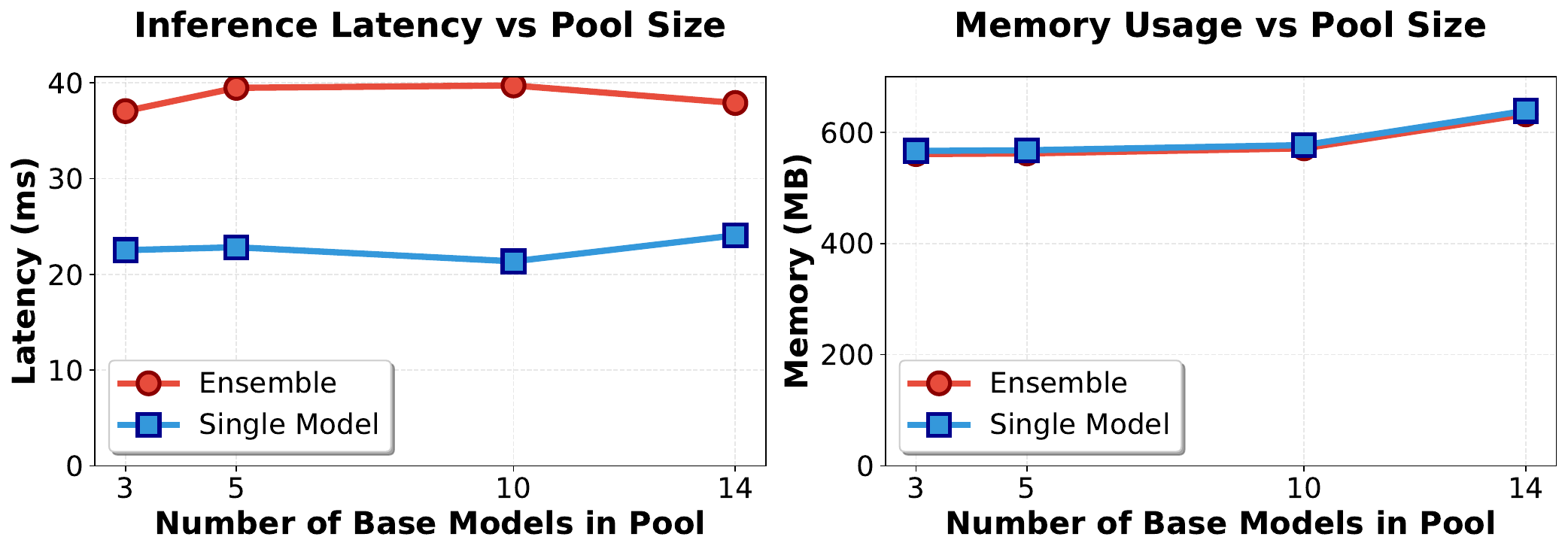}
\vspace{-7mm} 
\caption{Scalability: F1 and latency vs. candidate pool size for SKAB dataset.}
\label{fig:scalability_detectors}
\vspace{-2mm}
\end{figure}

Fig.~\ref{fig:scalability_detectors} shows runtime and memory consumption in relatio to the the candidate pool size. Both resource footprints are approximately constant from 3 to 14, which is expected as \system is only evaluating and using the models not retraining them in online phase for each window.

\subsubsection{Altering Online Window Size}
\label{subsubsection:window-size-effect}
\begin{figure}[ht]
\centering
\includegraphics[width=0.50\textwidth]{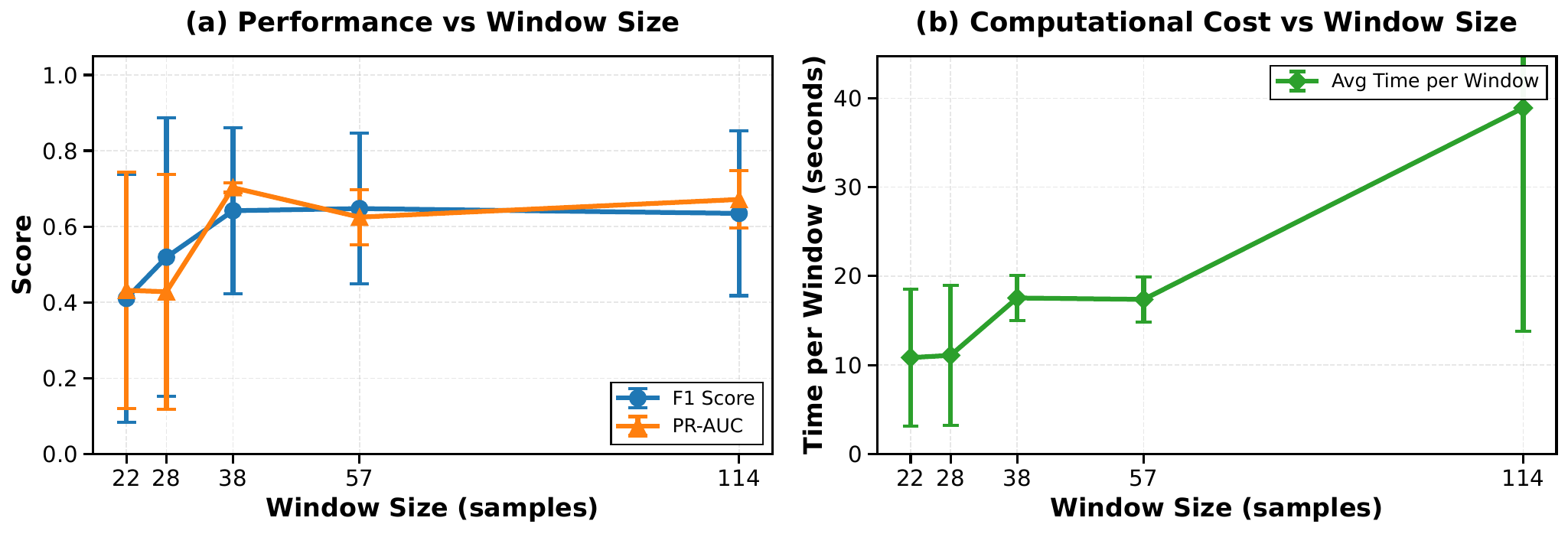}
\vspace{-7mm} 
\caption{Window size sensitivity on SKAB dataset.}
\label{fig:window_size_sensitivity}
\vspace{-4mm}
\end{figure}

Fig.~\ref{fig:window_size_sensitivity} examines average performance and latency vs. window sizes varying from $22$ to $114$ timesteps over three time series from SMD. \system performs poorly with smaller window sizes due to insufficient context fed to the detectors, and the detectors exhibit behavior similar to random guessing. Performance increases significantly as window size grows, reaching a plateau where further increases yield diminishing returns. Runtime overhead increases with larger window sizes because each window requires processing more samples for detection.

\subsection{Branch Performance Comparison}
\label{subsec:branch_comparison}

Table~\ref{tab:branch_performance} compares GA-Ens against single-model selection across datasets and domains, using the exact hyperparameter values from Table~\ref{tab:hyperparameters} fixed throughout all experiments.

\begin{table}[t]
\centering
\caption{Branch average performance comparison based on F1-score.}
\label{tab:branch_performance}
\scriptsize
\setlength{\tabcolsep}{3pt}
\renewcommand{\arraystretch}{1.15}
\begin{tabular}{@{}llcccc@{}}
\toprule
\textbf{Dataset} & \textbf{Domain} & \textbf{N} & \textbf{GA-Ens} & \textbf{Single Sel.} & \textbf{Final Sel.} \\
 & & & \textbf{F1±$\sigma$} & \textbf{F1±$\sigma$} & \textbf{F1±$\sigma$} \\
\midrule
UCR-Med & Healthcare & 68 & 0.77±0.06 & 0.80±0.09 & 0.82±0.07 \\
UCR-Env & Environment & 13 & 0.62±0.01 & 0.80±0.09 & 0.81±0.04 \\
UCR-Ind & Industry & 33 & 0.72±0.07 & 0.80±0.09 & 0.82±0.07 \\
\midrule
SMD & IT & 25 & 0.88±0.07 & 0.87±0.05 & 0.89±0.05 \\
\midrule
SKAB & Industry & 15 & 0.78±0.04 & 0.55±0.30 & 0.80±0.07 \\
\bottomrule
\end{tabular}
\vspace{-5mm}
\end{table}

\textit{Univariate vs. Multivariate.}
Table~\ref{tab:branch_performance} shows the single-branch achieves higher F1 across all univariate UCR subsets. With 1 feature, GA-Ens meta-learner lacks information about relevant cross-feature relationships. In contrast, single-model selection identifies detectors that robustly capture temporal patterns within the univariate stream through LinTS exploration and stress testing. The ensemble-branch excels on multivariate series but highly depends on the relationships between features. On SMD with 38 features both branches perform similarly. However on SKAB  with 8 features, the ensemble branch captures cross-sensor dependencies, e.g., valve failure manifesting across pressure and temperature simultaneously, that single-model selection cannot exploit.
On \textit{UCR-Medical} LinTS's evaluation captures regime shifts, e.g., arrhythmia onset, better than GA-Ens. GAN generates borderline anomalies that validate single-model generalization. On \textit{UCR-Environment:} Big single-branch F1 advantage =0.80 vs. GA-Ens=0.62. SBA, 10\% injection, scale [0.95,1.05]) matches temperature anomaly patterns (near-threshold spikes). On \textit{Industrial (SKAB):} Sensor correlations (valve failure → pressure drop + temperature rise) require ensemble. SKAB's poor single-branch performance stems from insufficient data for LinTS: $<$30 points per window, which is inadequate for stable covariance estimation.

\noindent\textit{Guideline for practitioners.}
Dimensionality and cross-feature dependencies determine the branch choice: ensemble methods leverage complex inter-feature relationships in multivariate settings, while single-model selection suffices for simple patterns with sufficient data points. Table~\ref{tab:branch_performance} confirms this, as single-branch excels on univariate UCR, GA-Ens excels on multivariate SMD/SKAB. However, we can also observe that single-branch takes longer on the long UCR series, while the ensemble-branch adds overhead for high-dimensional data, requiring practitioners to trade-off accuracy versus efficiency.
\textit{In summary,} (1) when accuracy is paramount, data is multivariate with inter-feature relationships, prefer ensemble-branch. (2) When latency is critical and data is univariate, prefer single-branch. (3) For unknown characteristics, deploy both branches and select adaptively based on online performance.

\subsection{Ensemble Branch Analysis}
\label{subsection:Ensemble Selection Results}
To better understand the ensemble branch, we investigate its hyperparameter effects and runtime efficiency.

\subsubsection{Effect of Population Size and Generations}

To evaluate the convergence of GA-Ens, we varied the \emph{P} and \emph{G} for $\mu=0.2$ and SVM as the meta-learner. Table~\ref{table:convergence} reports the resulting F1 scores.
For $P \in \{10,50,100\}$ and $G \in \{100,1000,10000\}$, both the chosen ensembles and their F1 remain nearly stable. For example, entities such as \texttt{SKAB\_1-1}, and \texttt{SMD\_3-10} consistently converge to the same or very similar sets of base models. This indicates that the GA reliably identifies strong ensemble configurations without extensive tuning of \emph{P} or \emph{G}.



\begin{table}[t]
\centering
\caption{F1 across $P$ sizes and $G$. Meta-learner= SVM and $\mu$= 0.2.}
\vspace{-.2cm}
\label{table:convergence}
\scriptsize
\setlength{\tabcolsep}{3pt}
\begin{tabular}{@{}l *{9}{c}@{}}
\toprule
Entity & 
\multicolumn{3}{c}{P 10} & 
\multicolumn{3}{c}{P 50} & 
\multicolumn{3}{c}{P 100} \\
\cmidrule(lr){2-4} \cmidrule(lr){5-7} \cmidrule(lr){8-10}
 & G 100 & 1000 & 10000 & 100 & 1000 & 10000 & 100 & 1000 & 10000 \\
\midrule
SKAB 1-1  & 0.87 & 0.87 & 0.87 & 0.87 & 0.87 & 0.87 & 0.87 & 0.87 & 0.87 \\
SKAB 1-2  & 0.87 & 0.87 & 087 & 0.87 & 0.87 & 0.87 & 0.87 & 0.87 & 0.87 \\
SMD 3-4   & 0.77 & 0.76 & 0.76 & 0.76 & 0.76 & 0.76 & 0.76 & 0.76 & 0.76 \\
SMD 3-10  & 0.71 & 0.71 & 0.71 & 0.71 & 0.71 & 0.71 & 0.71 & 0.71 & 0.71 \\
011\_UCR  & 0.77 & 0.77 & 0.77 & 0.77 & 0.77 & 0.77 & 0.77 & 0.77 & 0.77 \\
012\_UCR  & 0.79 & 0.78 & 0.78 & 0.78 & 0.78 & 0.78 & 0.78 & 0.78 & 0.78 \\
\bottomrule
\end{tabular}
\end{table}


\subsubsection{Effect of the Mutation Rate (\texorpdfstring{$\mu$}{mu})}
Table~\ref{table:mutationrate} reports the F1 for each setting.
To study the impact of $\mu$ in the GA-Ens process, we vary $\mu$ in $\{0,\,0.05,\,0.2,\,1.0\}$ while keeping the \emph{P} $=100$, \emph{G} $=1000$, and meta-learner $=$ RF, to assess the effect of mutation rate on heavy loading.
For all values of $\mu$, the resulting ensembles exhibit \emph{stable} F1 scores. For example, on \texttt{SKAB\_1-1} and \texttt{SKAB\_1-2}, F1 remains above $0.86$, for both, despite differences in ensemble composition.
A moderate $\mu$, e.g., $0.2$, sustains diversity and helps avoid premature convergence. High $\mu$ values, e.g., $1.0$, do not degrade performance in our study, suggesting that \emph{P} contains multiple viable configurations.

\begin{table}[t]
\centering
\scriptsize
\begin{minipage}[t]{0.48\columnwidth}
\centering
\caption{F1 vs. $\mu$. $P$ = 100, $G$ = 1000, Meta-learner = RF.}
\vspace{-.2cm}
\label{table:mutationrate}
\setlength{\tabcolsep}{3pt}
\begin{tabular}{@{}lcccc@{}}
\toprule
Entity & 0 & 0.05 & 0.2 & 1 \\
\midrule
SKAB 1-1  & 0.87 & 0.87 & 0.87 & 0.87 \\
SKAB 1-2  & 0.86 & 0.86 & 0.86 & 0.86 \\
SMD 3-4   & 0.74 & 0.75 & 0.74 & 0.75 \\
SMD 3-10  & 0.76 & 0.76 & 0.76 & 0.76 \\
011\_UCR  & 0.78 & 0.78 & 0.78 & 0.78 \\
012\_UCR  & 0.79 & 0.79 & 0.79 & 0.79 \\
\bottomrule
\end{tabular}
\end{minipage}%
\hfill
\begin{minipage}[t]{0.48\columnwidth}
\centering
\caption{F1 vs. meta-learners. P = 100, G = 1000, $\mu$ = 0.2).}
\vspace{-.2cm}
\label{table:metalearner}
\setlength{\tabcolsep}{4pt}
\begin{tabular}{@{}lccc@{}}
\toprule
Entity & SVM & RF & LR \\
\midrule
SKAB 1-1  & 0.87 & 0.87 & 0.78 \\
SKAB 1-2  & 0.87 & 0.86 & 0.79 \\
SMD 3-4   & 0.76 & 0.74 & 0.66 \\
SMD 3-10  & 0.71 & 0.76 & 0.63 \\
011\_UCR  & 0.77 & 0.78 & 0.61 \\
012\_UCR  & 0.78 & 0.79 & 0.62 \\
\bottomrule
\end{tabular}
\end{minipage}
\vspace{-4mm}
\end{table}


\subsubsection{Effect of the Meta-Learner}
Table~\ref{table:metalearner} summarizes the F1 for the GA-Ens process and meta-learner variants.
To assess the influence of the meta-learner in our \emph{GA-Ens}, we compare SVM, RF, and LR, for $P=100$, $G=1000$, and $\mu=0.2$.
Across nearly all entities, \textit{SVM} outperforms both RF and LR. For example, on \texttt{SKAB\_1-1} and \texttt{SKAB\_1-2}, SVM achieves an F1 score of $0.87$ versus RF at $0.86$ and LR at $0.78$, for \texttt{SKAB\_1-1} and $0.79$ for \texttt{SKAB\_1-2}. 


\subsection{Runtime and Convergence of \system Ensemble}
\label{sec:runtime}

\begin{figure*}[htbp]
\centering
\scriptsize
\includegraphics[
  width=\linewidth,
scale=0.8,
]{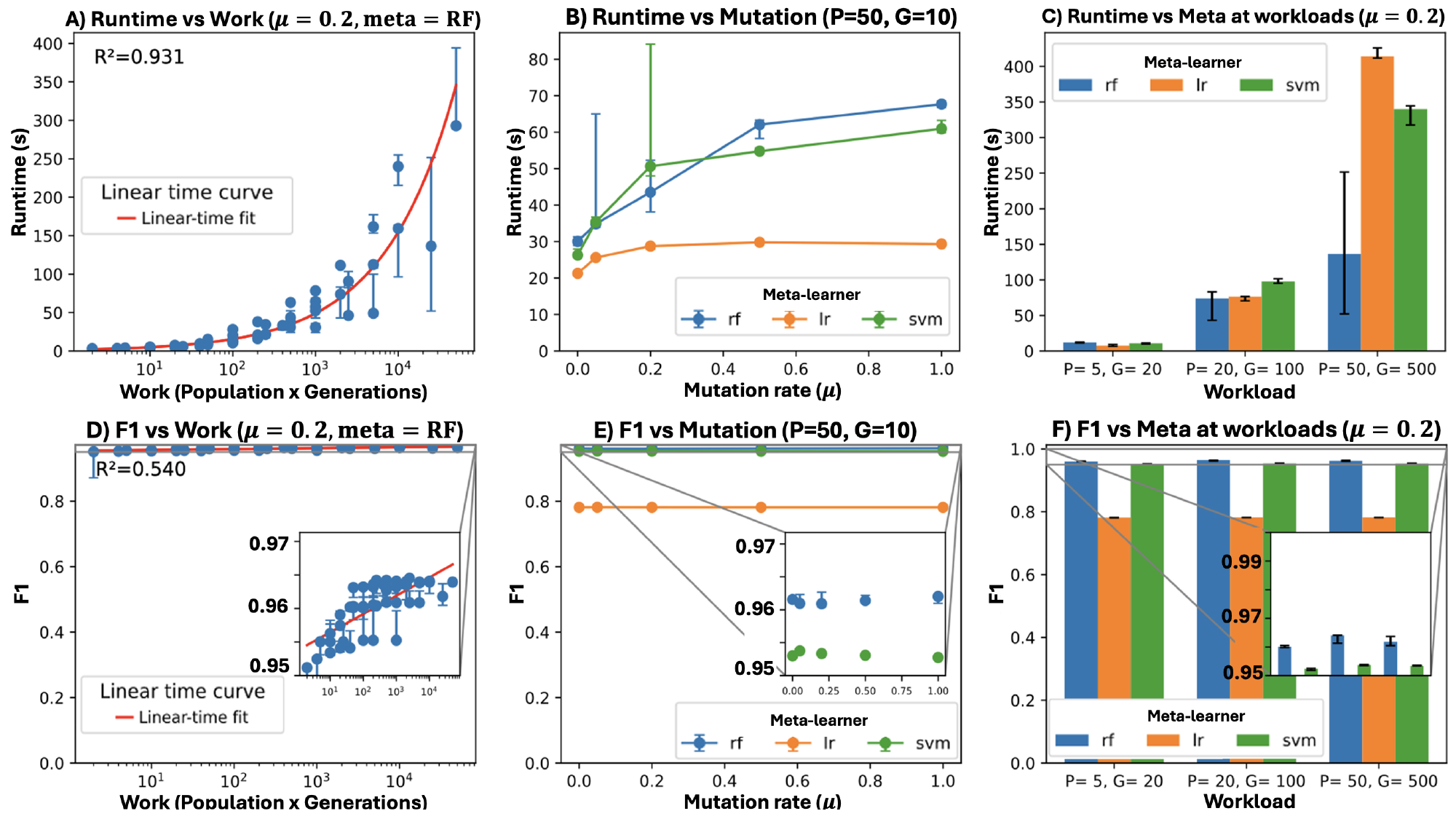}
\vspace{-7mm} 
\caption{Runtime–accuracy of the GA-based stacking ensemble on \texttt{SKAB1\_1}. (A) runtime vs.\ work ($P{\times}G$), $R^2{=}0.931$. (B) runtime vs.\ $\mu$ across meta-learners. (C) runtime vs.\ meta-learner at three workloads. (D) $F_1$ vs.\ work, $R^2{=}0.540$. (E) $F_1$ vs.\ $\mu$. (F) $F_1$ vs.\ meta-learner. Points are medians over 3 runs.}
\label{fig:ga_runtime_f1}
\vspace{-5mm} 
\end{figure*}

To analyze the runtime/accuracy trade-off of the \emph{GA-Ens} during the \emph{offline} selection stage we focus on \texttt{SKAB1\_1}. For each configuration, we ran three repeats and report medians with interquartile range. Wall-clock time includes the GA search and meta-learner training with cached base-model scores. The \emph{online} stage operates with the chosen ensemble and only triggers lightweight adaptation after accumulating user feedback as explained in~\ref{subsection:online}. Fig.~\ref{fig:ga_runtime_f1}A depicts runtime to grow linearly with \emph{work} ($P{\times}G$). A power-law fit describes the variance, as shown in Fig.~\ref{fig:ga_runtime_f1}A. $F_1$ reaches a plateau quickly as shown in Fig.~\ref{fig:ga_runtime_f1}D .
Figures.~\ref{fig:ga_runtime_f1}B, E depict the influence of $\mu$ on runtime and F1. At a fixed workload ($P{=}50$, $G{=}10$), runtime increases moderately with $\mu$ for RF/SVM, while LR remains nearly flat. $F_1$ is insensitive to $\mu$ for all meta-learners.
Across all workloads of $P$ and $G$, RF achieves near-top $F_1$ and is faster than SVM and LR as depicted in Figs.~\ref{fig:ga_runtime_f1}C, F, aligning with our results in Table~\ref{table:metalearner}.

\subsection{Single Model Selection Branch Analysis}




\noindent
To better understand behavior of single-model selection branch, Table~\ref{table:selection_methods} lists F1 score for four variants of our scoring module: \emph{LinTS}, \emph{GAN}, \emph{SBA}, and \emph{MC}. Across all entities, the selected models achieve nearly identical F1 and agree on the same model. This shows that the components are not contradicting, but supporting each other.


\begin{table}[ht]
\centering
\scriptsize
\begin{minipage}[t]{0.42\columnwidth}
\raggedright
\vspace{1mm}
\hspace{-0.9\columnwidth}
\caption{F1 of components.}
\vspace{-.2cm}
\label{table:selection_methods}
\centering
\setlength{\tabcolsep}{2pt}
\hspace{-0.3\columnwidth}
\begin{tabular}{@{}lcccc@{}}
\toprule
Entity & LinTS & GAN & SBA & MC \\
\midrule
SKAB 1-1  & 0.88 & 0.88 & 0.88 & 0.87 \\
SKAB 1-2  & 0.87 & 0.88 & 0.88 & 0.87 \\
SMD 3-4   & 0.90 & 0.86 & 0.90 & 0.90 \\
SMD 3-10  & 0.88 & 0.89 & 0.88 & 0.88 \\
011\_UCR  & 0.86 & 0.86 & 0.86 & 0.86 \\
012\_UCR  & 0.86 & 0.86 & 0.86 & 0.88 \\
\bottomrule
\end{tabular}
\end{minipage}%
\hspace{-0.04\columnwidth}%
\begin{minipage}[t]{0.55\columnwidth}
\centering
\caption{Rankings of components on SMD 3–10.}
\label{tab:markov_aggregation_smd310}
\vspace{-.2cm}
\setlength{\tabcolsep}{2pt}
\begin{tabular}{@{}cccccc@{}}
\toprule
GAN & SBA & MC & Rob. & LinTS & Final \\
\midrule
RNN4 & RNN3 & RNN3 & RNN3 & RNN3 & RNN3 \\
RNN3 & RNN4 & RNN4 & RNN4 & LOF4 & RNN1 \\
RNN2 & RNN2 & RNN2 & RNN2 & RNN1 & RNN4 \\
RNN1 & RNN1 & RNN1 & RNN1 & k-NN3  & RNN2 \\
LOF1 & LOF1 & LOF1 & LOF1 & RNN4 & LOF4 \\
LOF2 & LOF2 & LOF2 & LOF2 & k-NN1  & LOF1 \\
\bottomrule
\end{tabular}
\end{minipage}
\vspace{-5mm}
\end{table}
\subsection{Markov-Based Rank Aggregation}

Table~\ref{tab:markov_aggregation_smd310} reports the per-heuristic rankings, the intermediate aggregated rankings within the robustness group and  \emph{LinTS}, and the final ranking.
The results show strong agreement among the robustness-based methods, with models such as \texttt{RNN3} and \texttt{RNN4} consistently occupying the top positions. LinTS also ranks \texttt{RNN3} highly, while introducing a slight variance, e.g., \texttt{LOF4} and \texttt{RNN1} appear among the top-5 candidates. Markov aggregation over robustness signals yields an ordering favoring \texttt{RNN3}. Notably, \emph{final} aggregated ranking selects \texttt{RNN3}, \texttt{RNN1}, and \texttt{RNN4} as the top three models. This outcome aligns with the F1 results, confirming that the aggregation procedure integrates complementary signals.
\section{Limitations and Future Directions}
\label{sec:limitations}

We now point out several limitations of \system.

\textit{Memory Consumption at inference time.} Re-optimization during online phase introduces additional computational overhead. As depicted in Fig.~\ref{fig:online_phase_performance}, periodic re-execution of GA-Ens, LinTS and GAN incurs higher memory, limiting applicability in resource 
-constrained environments. Future work could adopt re-optimization triggered only when distribution shifts are detected or accuracy degradation.

\textit{Meta-Hyperparameter Configuration.} As described in Table~\ref{tab:hyperparameters}, \system introduces additional hyperparameters, which may limit the ease of deployment. Future work can automate tuning via transfer learning~\cite{DBLP:conf/icde/SuiWCXHZZYSP25_transferlearning}.

\textit{Dataset Scope.} Our evaluation focuses on univariate and multivariate numerical time series. \system's applicability to categorical, mixed-type, or event-based sequences remains unexplored and future work.

\textit{Interpretability.} The ensemble's meta-learner and Markov aggregation operate as black-box mechanisms. Future work could integrate explainability techniques such as SHAP~\cite{DBLP:conf/icdm/Takeishi19_shap, DBLP:conf/edbt/ShetiyaSA024_shapely} values or even LLMs to enhance transparency and trust.

\section{Conclusion}

We present \system, a novel framework to address the persistent challenges of model selection for TSAD. In comprehensive evaluations across diverse time-series, \system outperforms prior methods in both accuracy and adaptability.
A central strength of the proposed framework lies in its integration of ensemble learning with adaptive model optimization. By leveraging ensemble techniques in combination with genetic algorithms, \system intelligently selects and coordinates multiple base models, capturing diverse anomaly patterns.
Moreover, \system introduces a dynamic model selection mechanism based on Thompson sampling and $\varepsilon$-greedy reinforcement learning. This enables continuous policy refinement in response to shifting data distributions. 
To further ensure robustness, the framework incorporates adversarial and statistical validation through GAN, MC, and sensitivity analyses. These components assess the resilience of candidate models under noisy, imbalanced, or perturbed conditions providing robustness guarantees.

\section*{AI-Generated Content Acknowledgement}
We used ChatGPT (GPT-5) and Grammarly for grammar and spelling corrections, light stylistic polishing, and LaTeX table formatting. ChatGPT also suggested concise wording for figure/axis labels and helped create simple decorative icons for schematic illustrations. No technical content, datasets, analyses, or experimental results were generated by AI.
\bibliographystyle{IEEEtran}
\bibliography{abbreviations,ref}

\end{document}